\input harvmac

\input amssym
\input epsf


\newfam\frakfam
\font\teneufm=eufm10
\font\seveneufm=eufm7
\font\fiveeufm=eufm5
\textfont\frakfam=\teneufm
\scriptfont\frakfam=\seveneufm
\scriptscriptfont\frakfam=\fiveeufm


\def\bb{
\font\tenmsb=msbm10
\font\sevenmsb=msbm7
\font\fivemsb=msbm5
\textfont1=\tenmsb
\scriptfont1=\sevenmsb
\scriptscriptfont1=\fivemsb
}


\newfam\dsromfam
\font\tendsrom=dsrom10
\textfont\dsromfam=\tendsrom
\def\ds{\fam\dsromfam \tendsrom}


\newfam\mbffam
\font\tenmbf=cmmib10
\font\sevenmbf=cmmib7
\font\fivembf=cmmib5
\textfont\mbffam=\tenmbf
\scriptfont\mbffam=\sevenmbf
\scriptscriptfont\mbffam=\fivembf


\newfam\mbfcalfam
\font\tenmbfcal=cmbsy10
\font\sevenmbfcal=cmbsy7
\font\fivembfcal=cmbsy5
\textfont\mbfcalfam=\tenmbfcal
\scriptfont\mbfcalfam=\sevenmbfcal
\scriptscriptfont\mbfcalfam=\fivembfcal


\newfam\mscrfam
\font\tenmscr=rsfs10
\font\sevenmscr=rsfs7
\font\fivemscr=rsfs5
\textfont\mscrfam=\tenmscr
\scriptfont\mscrfam=\sevenmscr
\scriptscriptfont\mscrfam=\fivemscr

\def\<{\langle}
\def\>{\rangle}
\def\cO{{\cal O}}
\def\oo{\infty}
\def\de{\delta}
\def\De{\Delta}
\def\b{\beta}
\def\cE{{\cal E}}
\def\cS{{\cal S}}
\def\e{\epsilon}
\def\s{\sigma}



\def\frac#1#2{{#1 \over #2}}


\def\tilde{\widetilde}

\def\bar{\overline}
\def\bsq#1{{{\b{#1}}^{\lower 2.5pt\hbox{$\scriptstyle 2$}}}}
\def\bexp#1#2{{{\b{#1}}^{\lower 2.5pt\hbox{$\scriptstyle #2$}}}}
\def\dotexp#1#2{{{#1}^{\lower 2.5pt\hbox{$\scriptstyle #2$}}}}

\def\IL{\relax{\rm I\kern-.18em L}}
\def\IH{\relax{\rm I\kern-.18em H}}
\def\IR{\relax{\rm I\kern-.18em R}}
\def\IC{\relax{\rm I\kern-0.54 em C}}

\def\rt2{\sqrt{2}}
\def\half {{1 \over 2}}


\font\tenbifull=cmmib10
\font\tenbimed=cmmib7
\font\tenbismall=cmmib5
\textfont9=\tenbifull \scriptfont9=\tenbimed
\scriptscriptfont9=\tenbismall

\mathchardef\bbGamma="7000
\mathchardef\bbDelta="7001
\mathchardef\bbPhi="7002
\mathchardef\bbAlpha="7003
\mathchardef\bbXi="7004
\mathchardef\bbPi="7005
\mathchardef\bbSigma="7006
\mathchardef\bbUpsilon="7007
\mathchardef\bbTheta="7008
\mathchardef\bbPsi="7009
\mathchardef\bbOmega="700A
\mathchardef\bbalpha="710B
\mathchardef\bbbeta="710C
\mathchardef\bbgamma="710D
\mathchardef\bbdelta="710E
\mathchardef\bbepsilon="710F
\mathchardef\bbzeta="7110
\mathchardef\bbeta="7111
\mathchardef\bbtheta="7112
\mathchardef\bbkappa="7114
\mathchardef\bblambda="7115
\mathchardef\bbmu="7116
\mathchardef\bbnu="7117
\mathchardef\bbxi="7118
\mathchardef\bbpi="7119
\mathchardef\bbrho="711A
\mathchardef\bbsigma="711B
\mathchardef\bbtau="711C
\mathchardef\bbupsilon="711D
\mathchardef\bbphi="711E
\mathchardef\bbchi="711F
\mathchardef\bbpsi="7120
\mathchardef\bbomega="7121
\mathchardef\bbvarepsilon="7122
\mathchardef\bbvartheta="7123
\mathchardef\bbvarpi="7124
\mathchardef\bbvarrho="7125
\mathchardef\bbvarsigma="7126
\mathchardef\bbvarphi="7127

\def\IL{\relax{\rm I\kern-.18em L}}
\def\IH{\relax{\rm I\kern-.18em H}}
\def\IR{\relax{\rm I\kern-.18em R}}
\def\IC{\relax{\rm I\kern-0.54 em C}}





\def\CB{{\cal B}}

\def\CD{{\cal D}}

\def\CF{{\cal F}}

\def\CI{{\cal I}}

\def\CO{{\cal O}}

\def\CS{{\cal S}}


\def\1{{\ds 1}}
\def\R{\hbox{$\bb R$}}

\def\Z{\hbox{$\bb Z$}}


\def\ve{\varepsilon}

\noblackbox

\def\unit{\relax{\rm 1\kern-.26em I}}
\def\nada{\relax{\rm 0\kern-.30em l}}
\def\tilde{\widetilde}


\noblackbox
\def\IL{\relax{\rm I\kern-.18em L}}
\def\IH{\relax{\rm I\kern-.18em H}}
\def\IR{\relax{\rm I\kern-.18em R}}
\def\IC{\relax\hbox{$\inbar\kern-.3em{\rm C}$}}
\def\IZ{\relax\ifmmode\mathchoice
{\hbox{\cmss Z\kern-.4em Z}}{\hbox{\cmss Z\kern-.4em Z}} {\lower.9pt\hbox{\cmsss Z\kern-.4em Z}}
{\lower1.2pt\hbox{\cmsss Z\kern-.4em Z}}\else{\cmss Z\kern-.4em Z}\fi}

\def\CD {{\cal D}}
\def\CF {{\cal F}}

\def\partialslash{\not{\hbox{\kern-2pt $\partial$}}}

\def\CO {{\cal O}}

\def\CB {{\cal B}}

\def\CS {{\cal S}}


\def\cO{{\cal O}}
\def\CO {{\cal O}}

\def\CS {{\cal S }}

\font\manual=manfnt \def\dbend{\lower3.5pt\hbox{\manual\char127}}

\def\IZ{\relax\ifmmode\mathchoice
{\hbox{\cmss Z\kern-.4em Z}}{\hbox{\cmss Z\kern-.4em Z}} {\lower.9pt\hbox{\cmsss Z\kern-.4em Z}}
{\lower1.2pt\hbox{\cmsss Z\kern-.4em Z}}\else{\cmss Z\kern-.4em Z}\fi}
\def\half {{1\over 2}}

\def\bar{\overline}
\def\CS{{\cal S}}

\def\rt2{\sqrt{2}}
\def\irt2{{1\over\sqrt{2}}}

\def\slashchar#1{\setbox0=\hbox{$#1$}           
   \dimen0=\wd0                                 
   \setbox1=\hbox{/} \dimen1=\wd1               
   \ifdim\dimen0>\dimen1                        
      \rlap{\hbox to \dimen0{\hfil/\hfil}}      
      #1                                        
   \else                                        
      \rlap{\hbox to \dimen1{\hfil$#1$\hfil}}   
      /                                         
   \fi}

\def\foursqr#1#2{{\vcenter{\vbox{
    \hrule height.#2pt
    \hbox{\vrule width.#2pt height#1pt \kern#1pt
    \vrule width.#2pt}
    \hrule height.#2pt
    \hrule height.#2pt
    \hbox{\vrule width.#2pt height#1pt \kern#1pt
    \vrule width.#2pt}
    \hrule height.#2pt
        \hrule height.#2pt
    \hbox{\vrule width.#2pt height#1pt \kern#1pt
    \vrule width.#2pt}
    \hrule height.#2pt
        \hrule height.#2pt
    \hbox{\vrule width.#2pt height#1pt \kern#1pt
    \vrule width.#2pt}
    \hrule height.#2pt}}}}
\def\psqr#1#2{{\vcenter{\vbox{\hrule height.#2pt
    \hbox{\vrule width.#2pt height#1pt \kern#1pt
    \vrule width.#2pt}
    \hrule height.#2pt \hrule height.#2pt
    \hbox{\vrule width.#2pt height#1pt \kern#1pt
    \vrule width.#2pt}
    \hrule height.#2pt}}}}
\def\sqr#1#2{{\vcenter{\vbox{\hrule height.#2pt
    \hbox{\vrule width.#2pt height#1pt \kern#1pt
    \vrule width.#2pt}
    \hrule height.#2pt}}}}

\def\figin{\epsfcheck\figin}\def\figins{\epsfcheck\figins}
\def\epsfcheck{\ifx\epsfbox\UnDeFiNeD
\message{(NO epsf.tex, FIGURES WILL BE IGNORED)}
\gdef\figin##1{\vskip2in}\gdef\figins##1{\hskip.5in}
\else\message{(FIGURES WILL BE INCLUDED)}%
\gdef\figin##1{##1}\gdef\figins##1{##1}\fi}
\def\DefWarn#1{}
\def\figinsert{\goodbreak\midinsert}
\def\ifig#1#2#3{\DefWarn#1\xdef#1{fig.~\the\figno}
\writedef{#1\leftbracket fig.\noexpand~\the\figno}%
\figinsert\figin{\centerline{#3}}\medskip\centerline{\vbox{\baselineskip12pt \advance\hsize by
-1truein\noindent\footnotefont{\bf Fig.~\the\figno:\ } \it#2}}
\bigskip\endinsert\global\advance\figno by1}


\lref\ZamolodchikovGT{
  A.~B.~Zamolodchikov,
  ``Irreversibility of the Flux of the Renormalization Group in a 2D Field Theory,''
JETP Lett.\  {\bf 43}, 730 (1986), [Pisma Zh.\ Eksp.\ Teor.\ Fiz.\  {\bf 43}, 565 (1986)].
}

\lref\HartnollRZA{
  S.~A.~Hartnoll, D.~M.~Ramirez and J.~E.~Santos,
  ``Thermal conductivity at a disordered quantum critical point,''
[arXiv:1508.04435 [hep-th]].
}

\lref\OsbornGM{
  H.~Osborn,
  ``Weyl consistency conditions and a local renormalization group equation for general renormalizable field theories,''
Nucl.\ Phys.\ B {\bf 363}, 486 (1991).
}

\lref\RattazziPE{
  R.~Rattazzi, V.~S.~Rychkov, E.~Tonni and A.~Vichi,
  ``Bounding scalar operator dimensions in 4D CFT,''
JHEP {\bf 0812}, 031 (2008).
[arXiv:0807.0004 [hep-th]].
}

\lref\DiPietroTAA{
  L.~Di Pietro, Z.~Komargodski, I.~Shamir and E.~Stamou,
 ``Quantum Electrodynamics in d=3 from the epsilon-expansion,''
[arXiv:1508.06278 [hep-th]].
}

\lref\HartnollCUA{
  S.~A.~Hartnoll and J.~E.~Santos,
  ``Disordered horizons: Holography of randomly disordered fixed points,''
Phys. Rev.\ Lett.\  {\bf 112}, 231601 (2014).
[arXiv:1402.0872 [hep-th]].
}

\lref\HartnollFAA{
  S.~A.~Hartnoll, D.~M.~Ramirez and J.~E.~Santos,
  ``Emergent scale invariance of disordered horizons,''
[arXiv:1504.03324 [hep-th]].
}

\lref\CaselleCSA{
  M.~Caselle, G.~Costagliola and N.~Magnoli,
  ``Numerical determination of the operator-product-expansion coefficients in the 3D Ising model from off-critical correlators,''
Phys.\ Rev.\ D {\bf 91}, no. 6, 061901 (2015).
[arXiv:1501.04065 [hep-th]].
}

\lref\BoundaryRG{
M.~Pleimling et al., ``Logarithmic corrections in the two-dimensional Ising model in a random surface field.'' Journal of physics A: mathematical and general 37.37 (2004): 8801.}

\lref\BoundaryRGC{
J.~Cardy, ``The Ising model in a random boundary field.'' Journal of Physics A: Mathematical and General 24.22 (1991): L1315.}

\lref\Dotsenko{
  V.~Dotsenko, M.~Picco, and P.~Pujol. "Renormalisation-group calculation of correlation functions for the 2D random bond Ising and Potts models." Nuclear Physics B 455.3 (1995): 701-723.}
 
\lref\Gurarie{V.~Gurarie, ``c-Theorem for disordered systems," Nuclear Physics B 546.3 (1999): 765-778. [cond-mat/9808063].}

\lref\BrezinQA{
  E.~Brezin and J.~Zinn-Justin,
  ``Spontaneous Breakdown of Continuous Symmetries Near Two-Dimensions,''
Phys.\ Rev.\ B {\bf 14}, 3110 (1976).
}

\lref\LC{
A.~Ludwig and J.~Cardy. "Perturbative evaluation of the conformal anomaly at new critical points with applications to random systems." Nuclear Physics B 285 (1987): 687-718.	}

\lref\StressCardy{J.~Cardy, ``The stress tensor in quenched random systems," Statistical Field Theories 73 (2002): 215-222. [cond-mat/0111031].}

\lref\VV{
Vik.~S.~Dotsenko and Vl.~S.~Dotsenko, "Critical behaviour of the 2D Ising model with impurity bonds." Journal of Physics C: Solid State Physics 15.3 (1982): 495.}

\lref\Hertz{
J.~Hertz, ``Disordered systems," Physica Scripta 1985.T10 (1985): 1.}

\lref\CardyXT{
  J.~L.~Cardy,
  ``Scaling and renormalization in statistical physics,''
Cambridge, UK: Univ. Pr. (1996) 238 p. (Cambridge lecture notes in physics: 3).
}

\lref\Belanger{D.~P.~Belanger, ``Experimental characterization of the Ising model in disordered antiferromagnets." Brazilian Journal of Physics 30.4 (2000): 682-692.}

\lref\ElShowkHT{
  S.~El-Showk, M.~F.~Paulos, D.~Poland, S.~Rychkov, D.~Simmons-Duffin and A.~Vichi,
  ``Solving the 3D Ising Model with the Conformal Bootstrap,''
Phys.\ Rev.\ D {\bf 86}, 025022 (2012).
[arXiv:1203.6064 [hep-th]].
}

\lref\KosBKA{
  F.~Kos, D.~Poland and D.~Simmons-Duffin,
  ``Bootstrapping Mixed Correlators in the 3D Ising Model,''
JHEP {\bf 1411}, 109 (2014).
[arXiv:1406.4858 [hep-th]].
}

\lref\SimmonsDuffinQMA{
  D.~Simmons-Duffin,
  ``A Semidefinite Program Solver for the Conformal Bootstrap,''
JHEP {\bf 1506}, 174 (2015).
[arXiv:1502.02033 [hep-th]].
}

\lref\improvementunpublished{
  L.~Iliesiu, F.~Kos, D.~Poland, S.~S.~Pufu, D.~Simmons-Duffin and R.~Yacoby,
  ``(unpublished).''
}

\lref\KosTGA{
  F.~Kos, D.~Poland and D.~Simmons-Duffin,
  ``Bootstrapping the $O(N)$ vector models,''
JHEP {\bf 1406}, 091 (2014).
[arXiv:1307.6856 [hep-th]].
}

\lref\PolandEY{
  D.~Poland, D.~Simmons-Duffin and A.~Vichi,
  ``Carving Out the Space of 4D CFTs,''
JHEP {\bf 1205}, 110 (2012).
[arXiv:1109.5176 [hep-th]].
}

\lref\ONFuture{
 F.~Kos, D.~Poland, D.~Simmons-Duffin, and A.~Vichi,
 ``Precision Islands in the Ising and $O(n)$ Models,''
 {\it to appear.}
}

\lref\PappadopuloJK{
  D.~Pappadopulo, S.~Rychkov, J.~Espin and R.~Rattazzi,
  ``OPE Convergence in Conformal Field Theory,''
Phys.\ Rev.\ D {\bf 86}, 105043 (2012).
[arXiv:1208.6449 [hep-th]].
}

\lref\RychkovLCA{
  S.~Rychkov and P.~Yvernay,
  ``Remarks on the Convergence Properties of the Conformal Block Expansion,''
[arXiv:1510.08486 [hep-th]].
}

\lref\KimOCA{
  H.~Kim, P.~Kravchuk and H.~Ooguri,
  ``Reflections on Conformal Spectra,''
[arXiv:1510.08772 [hep-th]].
}

\lref\LinWCG{
  Y.~H.~Lin, S.~H.~Shao, D.~Simmons-Duffin, Y.~Wang and X.~Yin,
  ``{\cal N}=4 Superconformal Bootstrap of the K3 CFT,''
[arXiv:1511.04065 [hep-th]].
}

\lref\WF{
K.~Wilson and M.~Fisher, ``Critical exponents in 3.99 dimensions," Physical Review Letters 28.4 (1972): 240.
}

\lref\AdamsRJ{
  A.~Adams and S.~Yaida,
  ``Disordered Holographic Systems I: Functional Renormalization,''
[arXiv:1102.2892 [hep-th]].
}

\lref\AdamsYI{
  A.~Adams and S.~Yaida,
  ``Disordered holographic systems: Marginal relevance of imperfection,''
Phys.\ Rev.\ D {\bf 90}, no. 4, 046007 (2014).
[arXiv:1201.6366 [hep-th]].
}

\lref\PolandWG{
  D.~Poland and D.~Simmons-Duffin,
  ``Bounds on 4D Conformal and Superconformal Field Theories,''
JHEP {\bf 1105}, 017 (2011).
[arXiv:1009.2087 [hep-th]].
}

\lref\ElShowkHU{
  S.~El-Showk and M.~F.~Paulos,
  ``Bootstrapping Conformal Field Theories with the Extremal Functional Method,''
Phys.\ Rev.\ Lett.\  {\bf 111}, no. 24, 241601 (2013).
[arXiv:1211.2810 [hep-th]].
}

\lref\OsbornGM{
  H.~Osborn,
  ``Weyl consistency conditions and a local renormalization group equation for general renormalizable field theories,''
Nucl.\ Phys.\ B {\bf 363}, 486 (1991).
}

\lref\HartnollCUA{
  S.~A.~Hartnoll and J.~E.~Santos,
  ``Disordered horizons: Holography of randomly disordered fixed points,''
Phys. Rev.\ Lett.\  {\bf 112}, 231601 (2014).
[arXiv:1402.0872 [hep-th]].
}

\lref\HartnollFAA{
  S.~A.~Hartnoll, D.~M.~Ramirez and J.~E.~Santos,
  ``Emergent scale invariance of disordered horizons,''
[arXiv:1504.03324 [hep-th]].
}

\lref\CaselleCSA{
  M.~Caselle, G.~Costagliola and N.~Magnoli,
  ``Numerical determination of the operator-product-expansion coefficients in the 3D Ising model from off-critical correlators,''
Phys.\ Rev.\ D {\bf 91}, no. 6, 061901 (2015).
[arXiv:1501.04065 [hep-th]].
}

\lref\KomargodskiXV{
  Z.~Komargodski,
  ``The Constraints of Conformal Symmetry on RG Flows,''
JHEP {\bf 1207}, 069 (2012).
[arXiv:1112.4538 [hep-th]].
}

\lref\Gurarie{V.~Gurarie, ``c-Theorem for disordered systems," Nuclear Physics B 546.3 (1999): 765-778. [cond-mat/9808063].}

\lref\StressCardy{J.~Cardy, ``The stress tensor in quenched random systems," Statistical Field Theories 73 (2002): 215-222. [cond-mat/0111031].}

\lref\Hertz{
J.~Hertz, ``Disordered systems," Physica Scripta 1985.T10 (1985): 1.}

\lref\CardyXT{
  J.~L.~Cardy,
  ``Scaling and renormalization in statistical physics,''
Cambridge, UK: Univ. Pr. (1996) 238 p. (Cambridge lecture notes in physics: 3).
}

\lref\Belanger{D.~P.~Belanger, ``Experimental characterization of the Ising model in disordered antiferromagnets." Brazilian Journal of Physics 30.4 (2000): 682-692.}

\lref\ElShowkHT{
  S.~El-Showk, M.~F.~Paulos, D.~Poland, S.~Rychkov, D.~Simmons-Duffin and A.~Vichi,
  ``Solving the 3D Ising Model with the Conformal Bootstrap,''
Phys.\ Rev.\ D {\bf 86}, 025022 (2012).
[arXiv:1203.6064 [hep-th]].
}

\lref\FaulknerJY{
  T.~Faulkner, H.~Liu and M.~Rangamani,
  ``Integrating out geometry: Holographic Wilsonian RG and the membrane paradigm,''
JHEP {\bf 1108}, 051 (2011).
[arXiv:1010.4036 [hep-th]].
}

\lref\WF{
K.~Wilson and M.~Fisher, ``Critical exponents in 3.99 dimensions," Physical Review Letters 28.4 (1972): 240.
}

\lref\AdamsRJ{
  A.~Adams and S.~Yaida,
  ``Disordered Holographic Systems I: Functional Renormalization,''
[arXiv:1102.2892 [hep-th]].
}

\lref\AdamsYI{
  A.~Adams and S.~Yaida,
  ``Disordered holographic systems: Marginal relevance of imperfection,''
Phys.\ Rev.\ D {\bf 90}, no. 4, 046007 (2014).
[arXiv:1201.6366 [hep-th]].
}

\lref\OKeeffeAWA{
  D.~K.~O'Keeffe and A.~W.~Peet,
  ``Perturbatively charged holographic disorder,''
[arXiv:1504.03288 [hep-th]].
}

\lref\AreanOAA{
  D.~Arean, A.~Farahi, L.~A.~Pando Zayas, I.~S.~Landea and A.~Scardicchio,
  ``Holographic p-wave Superconductor with Disorder,''
[arXiv:1407.7526 [hep-th]].
}

\lref\KlebanovTB{
  I.~R.~Klebanov and E.~Witten,
  ``AdS / CFT correspondence and symmetry breaking,''
Nucl.\ Phys.\ B {\bf 556}, 89 (1999).
[hep-th/9905104].
}

\lref\IM{
Y.~Imry and S-K.~Ma,
  ``Random-field instability of the ordered state of continuous symmetry," Physical Review Letters 35.21 (1975): 1399.
}

\lref\AYS{
A.~Aharony, Y.~Imry, and S-K.~Ma, ``Lowering of dimensionality in phase transitions with random fields," Physical Review Letters 37.20 (1976): 1364.}

\lref\ElShowkDWA{
  S.~El-Showk, M.~F.~Paulos, D.~Poland, S.~Rychkov, D.~Simmons-Duffin and A.~Vichi,
  ``Solving the 3d Ising Model with the Conformal Bootstrap II. c-Minimization and Precise Critical Exponents,''
J.\ Stat.\ Phys.\  {\bf 157}, 869 (2014).
[arXiv:1403.4545 [hep-th]].
}

\lref\wittenunpub{E. Witten, unpublished.}

\lref\BardeenPM{
  W.~A.~Bardeen and B.~Zumino,
  ``Consistent and Covariant Anomalies in Gauge and Gravitational Theories,''
Nucl.\ Phys.\ B {\bf 244}, 421 (1984).
}

\lref\MC{
S.~Fan, W.~Xiong, W.~Yuan, and F.~Zhong, ``Critical behavior of a three-dimensional random-bond Ising model using finite-time scaling with extensive Monte Carlo renormalization-group method," Physical Review E 81.5 (2010): 051132.}

\lref\OsbornGM{
  H.~Osborn,
  ``Weyl consistency conditions and a local renormalization group equation for general renormalizable field theories,''
Nucl.\ Phys.\ B {\bf 363}, 486 (1991).
}

\lref\CallanSA{
  C.~G.~Callan, Jr. and J.~A.~Harvey,
  ``Anomalies and Fermion Zero Modes on Strings and Domain Walls,''
Nucl.\ Phys.\ B {\bf 250}, 427 (1985).
}

\lref\Harris{
  A.~B.~Harris,
  ``Effect of random defects on the critical behaviour of Ising models,''
Journal of Physics C: Solid State Physics 7.9 (1974): 1671.
}

\lref\DotsenkoSY{
  V.~Dotsenko, M.~Picco and P.~Pujol,
  ``Renormalization group calculation of correlation functions for the 2-d random bond Ising and Potts models,''
Nucl.\ Phys.\ B {\bf 455}, 701 (1995).
[hep-th/9501017].
}
\lref\DotsenkoIM{
  V.~Dotsenko, M.~Picco and P.~Pujol,
  ``Spin spin critical point correlation functions for the 2-D random bond Ising and Potts models,''
Phys.\ Lett.\ B {\bf 347}, 113 (1995).
[hep-th/9405003].
}

\lref\ShimadaDM{
  H.~Shimada,
  ``Disordered O(n) Loop Model and Coupled Conformal Field Theories,''
Nucl.\ Phys.\ B {\bf 820}, 707 (2009).
[arXiv:0903.3787 [cond-mat.dis-nn]].
}

\lref\Dotsenko{
V.~Dotsenko, 
``Introduction to the replica theory of disordered statistical systems,'' Cambridge University Press, 2005.}

\lref\GliozziYSA{
  F.~Gliozzi,
  ``More constraining conformal bootstrap,''
Phys.\ Rev.\ Lett.\  {\bf 111}, 161602 (2013).
[arXiv:1307.3111].
}

\lref\GliozziJSA{
  F.~Gliozzi and A.~Rago,
  ``Critical exponents of the 3d Ising and related models from Conformal Bootstrap,''
JHEP {\bf 1410}, 042 (2014).
[arXiv:1403.6003 [hep-th]].
}

\lref\CardyRQG{
  J.~Cardy,
  ``Logarithmic conformal field theories as limits of ordinary CFTs and some physical applications,''
J.\ Phys.\ A {\bf 46}, 494001 (2013).
[arXiv:1302.4279 [cond-mat.stat-mech]].
}

\lref\ShangZW{
  Y.~Shang,
  ``Correlation functions in the holographic replica method,''
JHEP {\bf 1212}, 120 (2012).
[arXiv:1210.2404 [hep-th]].
}

\lref\Taka{
  M.~Fujita, Y.~Hikida, S.~Ryu and T.~Takayanagi,
  ``Disordered Systems and the Replica Method in AdS/CFT,''
JHEP {\bf 0812}, 065 (2008).
[arXiv:0810.5394 [hep-th]].
}

\lref\Niarcho{
  E.~Kiritsis and V.~Niarchos,
  ``Interacting String Multi-verses and Holographic Instabilities of Massive Gravity,''
Nucl.\ Phys.\ B {\bf 812}, 488 (2009).
[arXiv:0808.3410 [hep-th]].
}

\lref\BernardAS{
  D.~Bernard,
  ``(Perturbed) conformal field theory applied to 2-D disordered systems: An Introduction,''
In *Cargese 1995, Low-dimensional applications of quantum field theory* 19-61.
[hep-th/9509137].
}

\lref\AharonyAEA{
  O.~Aharony, Z.~Komargodski and S.~Yankielowicz,
  ``Disorder in Large-N Theories,''
[arXiv:1509.02547 [hep-th]].
}

\lref\Cardy{
J.~Cardy, ``Scaling and renormalization in statistical physics,'' Vol. 5. Cambridge university press, 1996.
}

\lref\ColemanCI{
  S.~R.~Coleman,
  ``There are no Goldstone bosons in two-dimensions,''
Commun.\ Math.\ Phys.\  {\bf 31}, 259 (1973).
}

\lref\MW{
D.~Mermin, and H.~Wagner, ``Absence of ferromagnetism or antiferromagnetism in one-or two-dimensional isotropic Heisenberg models," Physical Review Letters 17.22 (1966): 1133.}

\lref\KosTGA{
  F.~Kos, D.~Poland and D.~Simmons-Duffin,
  ``Bootstrapping the $O(N)$ vector models,''
JHEP {\bf 1406}, 091 (2014).
[arXiv:1307.6856 [hep-th]].
}

\lref\OsbornGM{
  H.~Osborn,
  ``Weyl consistency conditions and a local renormalization group equation for general renormalizable field theories,''
Nucl.\ Phys.\ B {\bf 363}, 486 (1991).
}

\lref\YurovYU{
  V.~P.~Yurov and A.~B.~Zamolodchikov,
  ``Truncated Conformal Space Approach To Scaling Lee-yang Model,''
Int.\ J.\ Mod.\ Phys.\ A {\bf 5}, 3221 (1990)..
}

\lref\LudwigRK{
  A.~W.~W.~Ludwig,
  ``Critical Behavior of the Two-dimensional Random $Q$ State Potts Model by Expansion in ($Q^-$2),''
Nucl.\ Phys.\ B {\bf 285}, 97 (1987).
}

\lref\OKeeffeAWA{
  D.~K.~O'Keeffe and A.~W.~Peet,
  ``Perturbatively charged holographic disorder,''
[arXiv:1504.03288 [hep-th]].
}

\lref\CostagliolaIER{
  G.~Costagliola,
  ``OPE Coefficients of the 3D Ising model with a trapping potential,''
[arXiv:1511.02921 [hep-th]].
}

\lref\fiveloops{B.~N.~Shalaev, S.~A.~Antonenko, and A.~I.~Sokolov, ``Five-loop $\sqrt\epsilon$ expansions for random Ising model and marginal spin dimensionality for cubic systems," Physics Letters A 230.1 (1997): 105-110.}

\lref\SKMa{ S-K.~Ma, ``Modern Theory of Critical Phenomena,'' No. 46. Da Capo Press, 2000.}

\lref\AreanOAA{
  D.~Arean, A.~Farahi, L.~A.~Pando Zayas, I.~S.~Landea and A.~Scardicchio,
  ``Holographic p-wave Superconductor with Disorder,''
[arXiv:1407.7526 [hep-th]].
}

\lref\IM{
Y.~Imry and S-K.~Ma,
  ``Random-field instability of the ordered state of continuous symmetry," Phys.\ Rev.\ Letters\ {\bf 35}, 1399 (1975).
}

\lref\AYS{
A.~Aharony, Y.~Imry, and S-K.~Ma, ``Lowering of dimensionality in phase transitions with random fields," Phys.\ Rev.\  Lett.\ {\bf 37}, 1364 (1976).}

\lref\Grinstein{G.~Grinstein, ``Ferromagnetic phase transition in random field: the breakdown of scaling laws," Phys.\  Rev.\  Lett.\ {\bf 37}, 944 (1976).}

\lref\Young{A.~P.~Young, ``On the lowering of dimenionality in phase transitions with random fields,'' J.~Phys.~C\ {\bf 10}, L257 (1977).}

\lref\ParisiKA{
  G.~Parisi and N.~Sourlas,
  ``Random Magnetic Fields, Supersymmetry and Negative Dimensions,''
Phys.\ Rev.\ Lett.\  {\bf 43}, 744 (1979).
}

\lref\ElShowkDWA{
  S.~El-Showk, M.~F.~Paulos, D.~Poland, S.~Rychkov, D.~Simmons-Duffin and A.~Vichi,
  ``Solving the 3d Ising Model with the Conformal Bootstrap II. c-Minimization and Precise Critical Exponents,''
J.\ Stat.\ Phys.\  {\bf 157}, 869 (2014).
[arXiv:1403.4545 [hep-th]].
}

\lref\Folk{R.~Folk, Y.~Holovatch, and T.~Yavorskii, ``Critical exponents of a three-dimensional weakly diluted quenched Ising model." Physics-Uspekhi 46.2 (2003): 169-191.}

\lref\MC{
S.~Fan, W.~Xiong, W.~Yuan, and F.~Zhong, ``Critical behavior of a three-dimensional random-bond Ising model using finite-time scaling with extensive Monte Carlo renormalization-group method," Physical Review E 81.5 (2010): 051132.}

\lref\OsbornGM{
  H.~Osborn,
  ``Weyl consistency conditions and a local renormalization group equation for general renormalizable field theories,''
Nucl.\ Phys.\ B {\bf 363}, 486 (1991).
}

\lref\DotsenkoSY{
  V.~Dotsenko, M.~Picco and P.~Pujol,
  ``Renormalization group calculation of correlation functions for the 2-d random bond Ising and Potts models,''
Nucl.\ Phys.\ B {\bf 455}, 701 (1995).
[hep-th/9501017].
}
\lref\DotsenkoIM{
  V.~Dotsenko, M.~Picco and P.~Pujol,
  ``Spin spin critical point correlation functions for the 2-D random bond Ising and Potts models,''
Phys.\ Lett.\ B {\bf 347}, 113 (1995).
[hep-th/9405003].
}

\lref\ShimadaDM{
  H.~Shimada,
  ``Disordered O(n) Loop Model and Coupled Conformal Field Theories,''
Nucl.\ Phys.\ B {\bf 820}, 707 (2009).
[arXiv:0903.3787 [cond-mat.dis-nn]].
}

\lref\Dotsenko{
V.~Dotsenko, 
``Introduction to the replica theory of disordered statistical systems,'' Cambridge University Press, 2005.}

\lref\GliozziYSA{
  F.~Gliozzi,
  ``More constraining conformal bootstrap,''
Phys.\ Rev.\ Lett.\  {\bf 111}, 161602 (2013).
[arXiv:1307.3111].
}

\lref\ElShowkNIA{
  S.~El-Showk, M.~Paulos, D.~Poland, S.~Rychkov, D.~Simmons-Duffin and A.~Vichi,
  ``Conformal Field Theories in Fractional Dimensions,''
Phys.\ Rev.\ Lett.\  {\bf 112}, 141601 (2014).
[arXiv:1309.5089 [hep-th]].
}

\lref\CardyRQG{
  J.~Cardy,
  ``Logarithmic conformal field theories as limits of ordinary CFTs and some physical applications,''
J.\ Phys.\ A {\bf 46}, 494001 (2013).
[arXiv:1302.4279 [cond-mat.stat-mech]].
}

\lref\HL{
A.~B.~Harris  and T. C. Lubensky, ``Renormalization-group approach to the critical behaviour of random-spin models," Physical Review Letters 33.26 (1974): 1540.
	}

\lref\Lubensky{T. C.  Lubensky, ``Critical properties of random-spin models from the $\epsilon$ expansion," Physical Review B 11.9 (1975): 3573.}

\lref\Khmel{ D.~E.~Khmelnitskii, ``Phase-transition of 2nd kind in inhomogeneous bodies," Zhurnal Eksperementalnoi i Teoretichiskoi Fiziki 68.5 (1975): 1960-1968.
	}

\lref\ShangZW{
  Y.~Shang,
  ``Correlation functions in the holographic replica method,''
JHEP {\bf 1212}, 120 (2012).
[arXiv:1210.2404 [hep-th]].
}

\lref\Taka{
  M.~Fujita, Y.~Hikida, S.~Ryu and T.~Takayanagi,
  ``Disordered Systems and the Replica Method in AdS/CFT,''
JHEP {\bf 0812}, 065 (2008).
[arXiv:0810.5394 [hep-th]].
}

\lref\RychkovNAA{
  S.~Rychkov and Z.~M.~Tan,
  ``The $\epsilon$-expansion from conformal field theory,''
J.\ Phys.\ A {\bf 48}, no. 29, 29FT01 (2015).
[arXiv:1505.00963 [hep-th]].
}

\lref\DiPietroTAA{
  L.~Di Pietro, Z.~Komargodski, I.~Shamir and E.~Stamou,
  ``Quantum Electrodynamics in d=3 from the epsilon-expansion,''
[arXiv:1508.06278 [hep-th]].
}

\lref\SenDOA{
  K.~Sen and A.~Sinha,
  ``On critical exponents without Feynman diagrams,''
[arXiv:1510.07770 [hep-th]].
}

\lref\ChesterWAO{
  S.~M.~Chester, M.~Mezei, S.~S.~Pufu and I.~Yaakov,
  ``Monopole Operators from the $4-\epsilon$ Expansion,''
[arXiv:1511.07108 [hep-th]].
}

\lref\DiabSPB{
  K.~Diab, L.~Fei, S.~Giombi, I.~R.~Klebanov and G.~Tarnopolsky,
  ``On $C_J$ and $C_T$ in the Gross-Neveu and $O(N)$ Models,''
[arXiv:1601.07198 [hep-th]].
}

\lref\BashmakovPCG{
  V.~Bashmakov, M.~Bertolini, L.~Di Pietro and H.~Raj,
  ``Scalar Multiplet Recombination at Large N and Holography,''
[arXiv:1603.00387 [hep-th]].
}

\lref\Niarcho{
  E.~Kiritsis and V.~Niarchos,
  ``Interacting String Multi-verses and Holographic Instabilities of Massive Gravity,''
Nucl.\ Phys.\ B {\bf 812}, 488 (2009).
[arXiv:0808.3410 [hep-th]].
}

\lref\Cardy{
J.~Cardy, ``Scaling and renormalization in statistical physics,'' Vol. 5. Cambridge university press, 1996.
}

\lref\Ludwig{A.~Ludwig and J.~Cardy. ``Perturbative evaluation of the conformal anomaly at new critical points with applications to random systems." Nuclear Physics B 285 (1987): 687-718.}

\lref\BanksWHA{
  E.~Banks, A.~Donos and J.~P.~Gauntlett,
  ``Thermoelectric DC conductivities and Stokes flows on black hole horizons,''
[arXiv:1507.00234 [hep-th]].
}

\lref\MW{
D.~Mermin, and H.~Wagner, ``Absence of ferromagnetism or antiferromagnetism in one-or two-dimensional isotropic Heisenberg models," Physical Review Letters 17.22 (1966): 1133.}

\lref\HogervorstRTA{
  M.~Hogervorst, S.~Rychkov and B.~C.~van Rees,
  ``Truncated conformal space approach in d dimensions: A cheap alternative to lattice field theory?,''
Phys.\ Rev.\ D {\bf 91}, 025005 (2015).
[arXiv:1409.1581 [hep-th]].
}

\lref\HogervorstAKT{
  M.~Hogervorst, S.~Rychkov and B.~C.~van Rees,
  ``Unitarity violation at the Wilson-Fisher fixed point in 4-epsilon dimensions,''
[arXiv:1512.00013 [hep-th]].
}

\lref\MaldacenaJN{
  J.~Maldacena and A.~Zhiboedov,
  ``Constraining Conformal Field Theories with A Higher Spin Symmetry,''
J.\ Phys.\ A {\bf 46}, 214011 (2013).
[arXiv:1112.1016 [hep-th]].
}

\def\figcaption#1#2{\DefWarn#1\xdef#1{Figure~\noexpand\hyperref{}{figure}%
{\the\figno}{\the\figno}}\writedef{#1\leftbracket Figure\noexpand~\xfig#1}%
\medskip\centerline{{\footnotefont\bf Figure~\hyperdef\hypernoname{figure}{\the\figno}{\the\figno}:}  #2 \wrlabeL{#1=#1}}%
\global\advance\figno by1}



\rightline{TAUP-2999/15, WIS/04/15-AUG-DPPA}
\Title{}
{
\vbox{
\centerline{The Random-Bond Ising Model}
\vskip10pt
\centerline{in 2.01 and 3 Dimensions}
}
}
\centerline{Zohar Komargodski$^1$ and David Simmons-Duffin$^2$} 
\vskip15pt

\centerline{ {\it $^1$ Weizmann Institute of Science, Rehovot 76100, Israel}}
\centerline{ {\it $^2$ School of Natural Sciences, Institute for Advanced Study, Princeton, New Jersey 08540}}

\vskip20pt

\centerline{\bf Abstract}
\noindent   

We consider the Ising model between 2 and 4 dimensions perturbed by quenched disorder in the strength of the interaction between nearby spins. In the interval $2<d<4$ this disorder is a relevant perturbation that drives the system to a new fixed point of the renormalization group. At $d=2$ such disorder is marginally irrelevant and can be studied using conformal perturbation theory. Combining  conformal perturbation theory with recent results from the conformal bootstrap we  compute some scaling exponents in an expansion around $d=2$. If one trusts these computations also in $d=3$, one finds results consistent with experimental data and Monte Carlo simulations. In addition, we perform a direct uncontrolled computation in $d=3$ using new results for low-lying operator dimensions and OPE coefficients in the 3d Ising model. We compare these new methods with previous studies. Finally, we comment about the $O(2)$ model in $d=3$, where we predict a large logarithmic correction to the infrared scaling of disorder.

\Date{March 2016}


\listtoc \writetoc

\newsec{Disorder in the Ising Model}
\seclab\disorderinisingmodel

Imagine that the interaction strength between nearby spins in the Ising model, $J_{\langle ij\rangle}$, (where $\langle ij\rangle$ denotes that $i$ and $j$ are neighboring spins) is approximately given by some constant $J$, but the fluctuations around $J$ are sampled from a normal distribution with a fixed standard deviation much smaller than $J$ itself. We may say that the interaction strength is subject to a weak, quenched disorder. Even though the disorder is arbitrarily weak, it may have profound effects on the infrared behavior of the theory. In particular, at the phase transition, the critical exponents may change. This problem arises in a variety of physical applications, see for example~\Belanger\ for a review. Near the phase transition, one can study the problem using the tools of Euclidean quantum field theory. 

In the Landau-Ginzburg description of the Ising fixed point, the energy operator, $\epsilon(x)$, corresponds to tuning the temperature slightly away from the critical temperature. In other words, if we deform the action (i.e.\ Hamiltonian) at the fixed point as 
\eqn\actionperturb{S\rightarrow S+h \int d^dx\, \epsilon(x),}
then to first order $h\sim T-T_c$. In fact, we can promote $h$ to a source, $h(x)$, and imagine that $T-T_c$ varies in space $h(x)\sim T(x)-T_c$. One can alternatively think of $h(x)$ as corresponding to the deviation of the interaction strength from its mean $h(x)\sim J-J(x)$.

What we would like to study is the situation where  $h(x)$ is a random variable with Gaussian distribution of width $c$:
\eqn\Gaussian{\overline{h(x)h(y)}\sim c^2\delta(x-y).} To study the disordered system we can imagine computing the ordinary free energy $\CF[h(x)]$ 
\eqn\ordinaryfree{e^{-\CF[h(x)]}=\int [\CD\sigma(x)]e^{-S-\int d^dxh(x)\epsilon(x)} , }
where $[\CD\sigma(x)]$ denotes a path integral over configurations of Ising spins. 
The disordered free energy can be obtained by averaging over $h(x)$:
\eqn\disorderfree{\CF_D=\int[\CD h(x)]\CF[h(x)] e^{-{1\over 2c^2}\int d^dx h^2(x)    }.}
The critical properties of the disordered system can be read from the disordered free energy as usual.

Let us now review the Harris criterion~\Harris. From~\ordinaryfree\ we see that it is natural to assign to $h$ dimension $[h]=d-\Delta_\epsilon$. Then the dimension of the strength of disorder is obtained from~\disorderfree\  $[c^2]=d-2\Delta_\epsilon$. Therefore, if $d<2\Delta_\epsilon$ disorder decays in the infrared and we say that it is irrelevant; The critical properties of the phase transition are unaltered by disorder. If $d\geq 2\Delta_\epsilon$ then disorder is relevant or marginal and may induce a renormalization group flow to a new fixed point.

Recall that the heat-capacity exponent is defined as
\eqn\heatcapa{\alpha\equiv {d-2\Delta_\epsilon\over d-\Delta_\epsilon},}
and therefore disorder is irrelevant if $\alpha<0$, marginal if $\alpha=0$, and relevant if $\alpha>0$.
The case $\alpha=0$ is especially interesting as there are logarithms and a beta function for $c^2$ can be defined. Furthermore, one can imagine that in favourable circumstances 
one can use the tools of conformal perturbation theory in order to systematically compute critical exponents in an expansion in $\alpha$.

In the 3d Ising model we have $\Delta_\epsilon\sim 1.41$ and therefore $\alpha\sim 0.11>0$ which means that the disorder is relevant. However, in the 2d Ising model $\Delta_\epsilon=1$ and thus $\alpha=0$, so disorder is marginal. As we will review, it turns out that it is marginally irrelevant. The conformal bootstrap techniques~\ElShowkNIA\ show that in $2+\ve$ dimensions, for small enough $\ve$, 
\eqn\expandE{\Delta_\epsilon=1+0.3\ve+\CO(\ve^2),\qquad \alpha=0.4\ve+\CO(\ve^2).}   
In other words, the heat capacity is small and positive for $\varepsilon>0$. In combination with the fact that the disorder is marginally irrelevant in $d=2$, this means that there is a perturbatively accessible fixed point for small enough $\ve$. Our goal is to compute some of the simplest properties of this disordered fixed point and compare the naive extrapolation to $d=3$ with experimental and Monte Carlo data. We also attempt an uncontrolled expansion directly in $d=3$ using new results for low-lying operator dimensions and OPE coefficients in the 3d Ising model, computed with numerical bootstrap techniques.

Since $\alpha=0$  also holds true in $d=4$, one can study the random bond Ising model in an expansion around four dimensions~\refs{\HL,\Lubensky,\Khmel}. However, in $d=4$ there are two marginal operators, namely, the strength of disorder~\disorderfree\ and also the usual quartic interaction $\sigma^4$. The RG flow occurs in a two-dimensional space and one finds a certain degeneracy at one loop, namely, one combination remains undetermined. This renders the expansion an expansion in $\sqrt{d-4}$. It is somewhat ineffective, at least at low orders. For example, the sign of the heat capacity in the disordered fixed point evaluated from the first nontrivial order is incorrect.\foot{See, for example,~\fiveloops\  for higher order computations.} 
(In theories where disorder is relevant in the ultraviolet, one would generally expect that it would be irrelevant in the infrared. Hence, while in such theories $\alpha\geq 0$ in the ultraviolet, one generally expects that $\alpha\leq 0$ in the infrared. This is why the sign of the infrared heat-capacity exponent is crucial. The statement that we expect $\alpha\leq 0$ can be put on rigorous grounds, see~\SKMa.)

In $d=2$, there are no marginal operators in the Ising model other than disorder~\disorderfree\ and the expansion is thus expected to be rather effective.

Our strategy in expanding around $d=2$ is very similar to the standard epsilon expansion~\WF.\foot{Recently there was renewed activity and interest in the epsilon expansion, mostly motivated by potential applications for quantum phase transitions. See, for example,~\refs{\RychkovNAA\DiPietroTAA\SenDOA\ChesterWAO\DiabSPB-\BashmakovPCG} and references therein.} The main conceptual difference being that we are not expanding around a Gaussian point in the ultraviolet, rather, we are expanding around a nontrivial ultraviolet conformal field theory (which is, however, quite well understood).  The potential utility of such an expansion for the random bond model has been already foreseen in section 8 of~\Cardy.  For other related works see~\refs{\LudwigRK\DotsenkoIM\DotsenkoSY-\ShimadaDM}. In particular,~\LudwigRK\ studied the disordered $d=2$ Potts model by an expansion in the number of spin components.  The general framework of a controlled expansion around a nontrivial fixed point with a weakly relevant operator has had various applications. For example, the computation of central charges in such renormalization group flows (including in disordered theories) can be found in~\LC.

\newsec{The Replica Method }
\seclab\thereplicamethod

A popular approach to studying the disordered free energy~\disorderfree\ is via the replica method. In this note our main interest will be in theories where $c^2$ remains small under renormalization group transformations and one can thus attempt a series expansion in $c^2$. In this case, the replica method can be viewed simply as a convenient book-keeping device. The analytic continuation necessary in the replica method is guaranteed to exist.  

We define
\eqn\replica{W_n\equiv \int [{\cal D}h] e^{-n\CF[h(x)]} e^{-\int d^dx {h(x)h(x)\over 2c^2}},}
where $e^{-\CF[h(x)]}$ is defined in~\ordinaryfree.
We can compute the averaged disordered free energy from 
\eqn\continuation{\CF_D={d\over  dn} W_n\biggr|_{n=0}.}

In order to compute~\replica\ we need a convenient expression for $e^{-n\CF[h(x)]}$. For integer $n$, this can be achieved by introducing $n$ decoupled copies of the model 
\eqn\npower{e^{-n\CF[h(x)]}=\int \prod_{A=1}^n [{\cal D}\sigma_A]e^{-\sum_AS_A-\sum_A\int d^dx h(x)\epsilon_A(x)},  }
where capital Latin indices go over the replicas, $A=1,\dots,n$. We therefore have the path integral 
\eqn\npowerpi{W_n=\int [{\cal D}h] e^{-\int d^dx {h(x)h(x)\over 2c^2}} \prod_{A=1}^n [{\cal D}\sigma_A]e^{-\sum_AS_A-\sum_A\int d^dx h(x)\epsilon_A(x)}.}
We solve the $h$ path integral first. The equation of motion of $h$ sets $h(x)=-c^2\sum_A\epsilon_A(x)$ and thus
\eqn\npowerpi{W_n=\int\prod_{A=1}^n [{\cal D}\sigma_A]e^{-\sum_AS_A+{c^2\over 2}\int d^dx \sum_{A,B}\epsilon_A(x)\epsilon_B(x)}.}
Therefore, we have a collection of $n$ identical CFTs perturbed by some operator,  $ \sum_{A,B}\epsilon_A(x)\epsilon_B(x)$, with coupling $c^2$. 

The interaction $ \sum_{A,B}\epsilon_A(x)\epsilon_B(x)$ is relevant as long as $\Delta_\epsilon<d/2$ and marginal if $\Delta_\epsilon=d/2$. For the Ising$_d$ model, it is therefore relevant for $2<d<4$ and marginal if $d=2$ or $d=4$. See Fig~1.

\medskip
\vbox{
\offinterlineskip
\lineskip=3pt
\epsfxsize=3.0in \centerline{\epsfbox{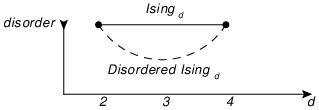}}
\nobreak\vskip0.4cm {
{\it \vbox{ \noindent {\bf Figure 1:}
{\it In $d=2,4$ disorder is marginal while in the interval $d\in(2,4)$ disorder is relevant and drives the theory to a new fixed point. (We will see that the situation in $d=4$ is actually more subtle than what the figure suggests.)}}}
}
}
\medskip

Let us consider the two marginal cases,  $d=2$ and $d=4$, in detail. 
One has to be careful about the case when the two indices in the sum $ \sum_{A,B}\epsilon_A(x)\epsilon_B(x)$ coincide, $A=B$. In $d=2$ the OPE of two energy operators does not contain marginal operators and thus one can throw away the terms $A=B$ (they only renormalize the energy of the ground state and do not affect the anomalous dimensions). However, in $d=4$ the OPE of two energy operators ($\epsilon={1\over \sqrt2 }\phi^2$, $\phi(x)\phi(0)\sim {1\over x^2}+\dots$) takes the form
\eqn\fourd{\epsilon(x)\epsilon(0)\sim {1\over x^4}+{2\sqrt 2\over x^2}\epsilon(0)+\epsilon^2(0)+\dots,}
where descendants have not been included in the OPE above.  
Therefore the terms with $A=B$ can be taken to stand for the $\epsilon^2(0)$ operator. Since the operator $\sum_A \epsilon^2_A$ is consistent with the permutation symmetry between the replicas, we have to add it to the action.

To summarize, in the two marginal cases corresponding to $d=2$ and $d=4$, we have to study the following theories 

\medskip

\item{d=2:} \eqn\dtwo{W_n=\int\prod_{A=1}^n [{\cal D}\sigma_A]e^{-\sum_AS_A+{c^2\over 2}\int d^2x \sum_{A\neq B}\epsilon_A(x)\epsilon_B(x)}.}
\item{d=4:} \eqn\dfour{W_n=\int\prod_{A=1}^n [{\cal D}\sigma_A]e^{-\sum_A\left(S_A+\lambda^2\int d^4x\  \epsilon^2_A(x)\right)+{c^2\over 2}\int d^4x \sum_{A,B}\epsilon_A(x)\epsilon_B(x)},}
where in the interaction term between the replicas, the terms with $A=B$ are interpreted simply as $\epsilon^2_A(x)$. We have included the coupling $\lambda^2$ even though it is  marginally irrelevant in the four-dimensional pure Ising model for reasons that will be clarified below. 

\medskip
We can study the $d=4$ theory using ordinary Feynman diagrams, but the theory in $d=2$ must be studied using conformal perturbation theory. We therefore begin with an extensive review of some aspects of conformal perturbation theory that will be useful in what follows.

\newsec{Review of Conformal Perturbation Theory}
\seclab\reviewofconformalperturbationtheory

Consider an operator $\CO$ with dimension $\Delta_\CO=d-\delta$. We will be interested in the case where $\CO$ is marginal, corresponding to $\delta=0$, and also the case where $\cO$ is slightly relevant, $0<\delta\ll 1$.  We assume that its OPE takes the form
\eqn\cpt{\CO(x)\CO(0)\sim {1\over x^{2\Delta_\CO}} +\frac{C_{\CO\CO\CO}}{x^{\Delta_\CO}}\cO(0)+\dots.}
Of course, the OPE may also contain various relevant operators, and it must  contain the energy-momentum tensor. We will comment on the effects of these terms later.

Imagine deforming the partition function by 
\eqn\marginalvev{ e^{g \int d^dx\, \CO(x)}.}
We can determine the beta function for $g$ by demanding that a physical observable be UV-cutoff independent.  A simple observable to compute is the overlap between the state $\<\cO|$ given by an insertion of 
\eqn\opatinfinity{\cO(\infty)\equiv \lim_{x\to \oo} x^{2\Delta_\CO}\CO(x),}
and the state $|0\>_{g,V}$ given by deforming the conformal theory with \marginalvev\ in a region of volume $V$ around the origin (we abuse notation and write $V$ for both the region and its volume),
\eqn\deformedvacuum{
|0\>_{g,V} \equiv e^{g\int_V d^d x\cO(x)}|0\>.
}
The restriction of the integral to $V$ ensures that the various integrated operators stay away from the insertion at infinity. In this way we are studying the renormalization of the unit operator, i.e.\ the free energy. 

Expanding in $g$ we find 
\eqn\conformalpertoverlap{\eqalign{
\<\CO| 0\>_{g,V} &= \langle \CO(\infty) \rangle+g\int_V d^dx\,\langle \CO(x) \CO(\infty)\rangle+\half g^2\int_V d^dx\,d^dx_1\langle \CO(x)\CO(x_1) \CO(\infty)\rangle \cr
&~~+{1\over 6} g^3\int_V d^dx\,d^dx_1\,d^dx_2\langle \CO(x)\CO(x_1)\CO(x_2) \CO(\infty)\rangle +\dots.
}}
Since we are interested in short distance divergences, we can set $x=0$ and replace $\int_V d^d x$ with a volume factor $V$, ignoring $x$-dependent boundary effects in the integrals over $x_{1,2}$. Then we have 
\eqn\freeenexp{\eqalign{
\<\CO| 0\>_{g,V} &\sim V g+\half  V g^2\int_V d^dx_1\langle \CO(0)\CO(x_1) \CO(\infty)\rangle\cr
&~~+{1\over 6} V g^3\int_V d^dx_1d^dx_2\langle \CO(0)\CO(x_1)\CO(x_2) \CO(\infty)\rangle +\dots,
}}
where ``$\sim$" means that both sides have the same UV-divergences.
Here we have used that the one-point function $\<\cO(\oo)\>$ vanishes and that $\<\cO(0)\cO(\infty)\>=1$.

The beta function up to two-loop order can be computed from divergences in the three-point and four-point functions in~\freeenexp.  For later convenience, we review this using two different regulators.  First, we consider the marginal case $\Delta_\CO=d$ with a cutoff regulator at the scale $\mu$.  Second, we consider the slightly relevant case $\Delta_\CO=d-\de$, $\de\ll 1$, using minimal subtraction of poles in $\de$.

\subsec{Marginal Case with Cutoff Regulator}

Suppose that $\cO$ is marginal, $\Delta(\cO)=d$.
The expansion~\freeenexp\ has logarithms depending on the cutoff. We absorb them by assuming that the effective coupling evolves with scale as $g(\mu)$. We demand that under a renormalization group transformation  the logarithms sum up in such a way that the answer is $\mu$-independent. The beta function is defined as usual \eqn\cseqs{\eqalign{  & {d\over d\log\mu}g(\mu)=\beta(g(\mu))=\beta_1g^2+\beta_2g^3+\dots.}}

Let us assume the integrated correlation functions in~\freeenexp\ diverge as 
\eqn\divone{\int_V d^dx_1\langle \CO(0)\CO(x_1) \CO(\infty)\rangle \sim A \log \mu,}
\eqn\divtwo{\int_V d^dx_1d^dx_2\langle \CO(0)\CO(x_1)\CO(x_2) \CO(\infty)\rangle \sim B\log^2\mu+C\log\mu.}
Then the Callan-Symanzik equation is 
\eqn\CS{{d\over d\mu}\left(g(\mu)+\half g^2(\mu) A\log\mu+{1\over 6}g^3(\mu)\left(B\log^2\mu+C\log\mu\right)\right)=0.}
Which allows us to obtain the equations
\eqn\CScons{\beta_1=-\half A,\qquad \beta_2=-{1\over 6 }C ,\qquad A^2 ={2\over 3}  B, }
with the last equation being the usual consistency condition that relates the logarithmic divergences at one and two loops. We clearly have $A=S_{d-1}C_{\CO\CO\CO}$, where $S_{d-1}={\rm Vol}(S^{d-1})={2\pi^{d/2}\over \Gamma(d/2)}$ is the volume of the unit $(d-1)$-sphere, and hence
\eqn\betaone{\beta(g)=-\half S_{d-1} C_{\CO\CO\CO}g^2+\dots.}

Let us explain why  $A^2={2\over 3}B$ is satisfied.
In~\divtwo\ we have three regions of divergences, i.e.~from the three points $0,x_1,x_2$ we can choose which pair is the closest in distance. If $|x_1|\ll |x_2|<1$ then the OPE in the channel $|x_1|\to 0$ will give an integral $C_{\CO\CO\CO}\int d^dx_1/x_1^d$ which needs to be cutoff at $|x_2|$ and thus we get 
$S_{d-1}C_{\CO\CO\CO}\log(\mu |x_2|)$. The remaining integral over $x_2$ is then $S_{d-1}C_{\CO\CO\CO}^2\int d^dx_2\log(\mu |x_2|)/x_2^d =\half S_{d-1}^2C_{\CO\CO\CO}^2 \log^2\mu$. 
Therefore, $B={3\over 2}S_{d-1}^2C_{\CO\CO\CO}^2$, as required by the Callan-Symanzik equation.

It remains to give a convenient expression for $C$, from which the beta function at two-loops can be extracted.

The integrated four-point function $\int d^dx_1d^dx_2\langle \CO(0)\CO(x_1)\CO(x_2) \CO(\infty)\rangle$ can be simplified since the integrand depends only on conformal cross-ratios. 
We have the two cross ratios 
\eqn\crossratios{u={|x_2|^2\over |x_1|^2 } ,\qquad v = {|x_1-x_2|^2\over |x_1|^2 }. }
The four-point function can be written as 
\eqn\fourptwritten{\langle \CO(0)\CO(x_1)\CO(x_2)\CO(\infty) \rangle={1\over |x_1|^{2d}}F\left({|x_2|\over |x_1| },{|x_1-x_2|\over |x_1| }\right).}
We write  the integration measure in spherical coordinates,  \eqn\sphericalcoords{d^dx_1d^dx_2= S_{d-1}S_{d-2}|x_{1}|^{d-1}|x_{2}|^{d-1}\sin^{d-2}(\Theta)d|x_{1}|d|x_{2}| d\Theta,} with $\Theta$ being the angle between $x_1$ and $x_2$. 

It is convenient to perform a change of variables $x_2\rightarrow x_2 |x_1|$ which brings the integral to the form 
\eqn\formi{S_{d-1} \int {d|x_1| \over |x_1|} \int d^d x_{2}  F\left(x_{2}  \right).} 
Henceforth, we may redefine $F$ to be the connected four-point function, since the disconnected piece does not contribute logarithmic divergences: 
\eqn\defineconnected{\langle \CO(0)\CO(e_1)\CO(x_2)\CO(\infty)\rangle_c =F(x_2),}
with $e_1=(1,0,0,\dots)$ a unit vector in the $1$-direction.
The expression~\formi\ is divergent. Indeed, it has the double logarithmic divergences that follow from the one-loop beta function according to~\CScons.\foot{It is instructive to re-derive the coefficient of the double-log from the expression~\formi. There are three regions with a $C^2_{\CO\CO\CO}/x^d$ singularity, namely, $x_2\rightarrow\{0,1,\infty\}$. But the cutoff in the $x_2$ coordinate is rescaled by $x_1$ because of the change of variables. Therefore we get the integral $3C_{\CO\CO\CO}^2\int {d|x_1| \over |x_1|} \log(|x_1|)\sim {3C_{\CO\CO\CO}^2\over 2} \log^2(\mu)$, which precisely agrees with~\CScons.  } The double-log needs to be carefully subtracted in order to compute $C$ in~\CS.

A simple trick to subtract the double-log is the following. One can consider generalized free field theory, deformed by a double trace operator. In this theory the beta function is one-loop exact. Therefore, one can subtract the double-log without affecting the single-log. This procedure respects all the symmetries. The derivation is in appendix A. This leads to the following expression 
\eqn\betatwo{\beta= -\half S_{d-1} C_{\CO\CO\CO}g^2  - {1\over 6}\int d^dx \left(F(x)- \half C^2_{\CO\CO\CO}  \left({1\over x^d (x-e_1)^d} +{1\over x^d}+{1\over (x-e_1)^d}\right)\right) g^3+\dots.}
The coefficient of $g^3$ is now a manifestly convergent integral.\foot{This is unless there are relevant operators in the OPE of $\CO(x)\CO(0)$ (other than the unit operator). Let us consider one such, normalized ($\langle\Psi(0)\Psi(\infty)\rangle=1$), relevant operator $\Psi$ of dimension $\Delta_\Psi<d$ such that the coefficient $C_\Psi$, defined as 
\eqn\definedas{\CO(x)\CO(0)\sim \frac{C_\Psi}{x^{2d-\Delta_\Psi}} \Psi(0)}
is non-vanishing.  This leads to a power divergence in~\betatwo.  We have to tune this away. For example, in the region that $x_2\rightarrow 0$ we simply subtract 
\eqn\simplysubtract{C_\Psi^2\int{d\rho\over \rho}\int d^d x_2 {1\over x_2^{2d-\Delta_\Psi}}.}
More generally,~\betatwo\ is modified to
\eqn\betatwoimp{\eqalign{\beta_2 =&{-1\over 6}\int d^dx \biggl[F(x)- \half C^2_{\CO\CO\CO}  \left({1\over x^d (x-e_1)^d} +{1\over x^d}+{1\over (x-e_1)^d}\right)\cr
&\qquad\qquad\qquad-C_\Psi^2\left({1\over x^{2d-\Delta_\psi}}+{1\over (x-e_1)^{2d-\Delta_\psi}}+x^{-\Delta_\psi}\right)\biggr].}}
If we sum over all the relevant operators, the integral~\betatwoimp\ becomes convergent. Note that the OPE of $\CO(x)\CO(0)$ necessarily contains the stress tensor, but this leads to no divergence in the integral~\betatwo\ due to the angular dependence of the leading OPE term.}

Finally, let us explain in what sense the procedure leading to~\betatwo\ is indeed unique. 
Different schemes correspond to contact terms. And since we are dealing here with integrated correlation functions, \divone, \divtwo, $\delta$-function singularities may contribute. We can for instance imagine introducing the contact term 
\eqn\contactexample{\cO(x)\cO(y)\sim \alpha\,\delta^{(d)}(x-y)\cO(x),}
with $\alpha$ some arbitrary coefficient. 
 From~\freeenexp\ we see that this amounts to a change of variables $g\rightarrow g+\half \alpha g^2$. However, it is well known that the first two terms in the beta function are invariant under this change of variables.\foot{Indeed, assume 
\eqn\indeedassume{\beta_g=\beta_1 g^2+\beta_2g^3+\dots.}
Now perform an analytic change of variables 
$g'=g+\half \alpha g^2+\dots$.
It turns out that $\alpha$ cancels and we have
\eqn\cancelsandwehave{\beta_{g'}=\beta_1 g'^2+\beta_2 g'^3+\dots.}
} Therefore,~\betatwo\ is scheme independent.

\subsec{Slightly Relevant Case}

We will also need to understand the case where $\cO$ is slightly relevant ($\De_\cO=d-\de$ with $0<\de\ll 1$).  There are two scenarios that one might consider:

\medskip

\item{1.} We have a family of CFTs where $\de$ is a continuous parameter that can be dialed to zero.  Assuming the CFT data has an analytic power-series expansion in $\de$, we can then regulate the deformed theory by minimal subtraction.  Specifically, we compute integrals like \freeenexp\ as a Laurent series in $\de$, and then add counterterms to cancel the poles.  Physical observables (like IR scaling dimensions) will be scheme-independent, and in principle can be computed to arbitrary order using these techniques.  An example of this situation is the usual Wilson-Fisher fixed point in $4-\epsilon$ dimensions. An example where the spacetime dimension does not vary is the Potts model where $q$ is regarded as a continuous parameter.

\item{2.} We might have an isolated CFT where $\de$ just happens to be numerically small.  In this case, the Laurent expansion of integrals like  \freeenexp\ becomes ambiguous beyond the leading nonzero order.  For example, without an expansion $C_{\cO\cO\cO}(\de)=C_{\cO\cO\cO}(0)+\de C'_{\cO\cO\cO}(0)+\dots$~, we cannot compute subleading terms in expressions involving $C_{\cO\cO\cO}$.  In this scenario, minimal subtraction does not work. In fact, there is no scheme-independent expression for subleading corrections to physical observables in conformal perturbation theory.  This is ultimately because the IR theory lives at the end of a nontrivial RG flow. To learn about it, we must perform a nonperturbative diagonalization of the full Hamiltonian, perhaps using conformal truncation methods~\YurovYU\ (see also more recent work~\HogervorstRTA).  An example of this situation is the disordered 3d Ising model, where $\de\approx 0.18$. We will only be able to compute the leading nonzero contributions to disordered observables as an expansion in $\de$. In some cases these contributions occur at one loop, and in others at two loops.\foot{As we explain in section~8, the disordered 3d Ising model can be also viewed as a perturbation around (very close to) the disordered $O(2)$ model, thereby making it controllable. }

\medskip

For the moment, let us assume that scenario (1) is valid, and discuss how to compute quantities with minimal subtraction. Subsequently, we discuss which quantities are scheme-independent in scenario (2).

The nonzero value of $\de$ serves as a UV regulator for the integrals that appear in conformal perturbation theory.  Finite results in the limit $\de\to 0$ can be obtained by subtracting poles in $\de$.  Specifically, we deform the theory with a bare coupling
\eqn\barecoupling{
g_0 = \mu^\de \left(g + g^2{a \over \de} + g^3\left({b \over \de^2} + {c \over \de}\right) + \dots\right),
}
where the counterterms are chosen to subtract poles in $\de$.  Since the regulator is independent of $\mu$, physical observables will be $\mu$-independent if $g_0$ itself is $\mu$-independent,
\eqn\msbetafunction{
0 = {dg_0 \over d\log\mu}.
}
This gives the $\beta$ function
\eqn\msbetafunctiontwo{\eqalign{
{dg \over d \log \mu} &= -\de g + \b_1 g^2 + \b_2 g^3 + \dots \cr
&= -\de g + a g^2 + 2c g^3 + \dots,
}}
together with the relation between counterterms $b = a^2$.  

Plugging the bare coupling $g_0$ into the expansion \freeenexp\ for $\<\cO|0\>_{g_0,V}$, and choosing $a,b,c$ to cancel UV-divergences using minimal subtraction, we obtain
\eqn\msbetacoefficients{\eqalign{
\b_1 &= -{ 1\over 2}\left({1\over V}\int_V d^d x_1\, d^d x_2 \<\cO(x_1)\cO(x_2)\cO(\oo)\>\right)_{\de^{-1}},\cr
\b_2 &= -{ 1 \over 3}\left({1\over V}\int_V d^d x_1\,d^d x_2\, d^d x_3 \<\cO(x_1)\cO(x_2)\cO(x_3)\cO(\oo)\>\right)_{\de^{-1}},
}}
where $(\dots)_{\de^{-1}}$ means the term proportional to $\de^{-1}$ in a Laurent expansion in $\de$.

Let us now give a more precise prescription for computing these quantities.  We should find that the results agree with the expressions for $\b_k$ found in the previous subsection.  From the first integral, we easily obtain $\beta_1=-{1\over 2} S_{d-1} C_{\cO\cO\cO}$ as before. 
Note that here $C_{\cO\cO\cO}$ stands for the OPE coefficient evaluated with $\delta=0$. (This OPE coefficient, and, more generally, all the other CFT data are allowed to depend on $\delta$ smoothly.)

 For the second integral, it is convenient to isolate divergences by breaking the integration into three regions ${\cal R}_{12}$, ${\cal R}_{23}$, ${\cal R}_{31}$ given by
\eqn\integralregions{
{\cal R}_{12}: |x_{12}|<|x_{13}|,|x_{23}|
}
and cyclic permutations.  Each region contributes equally, giving a factor of $3$.  Within ${\cal R}_{12}$, we may cancel the factor of $V$ and set $x_1=0$.  We then use rotational invariance to set $x_3=r e_1$, giving
\eqn\betaonems{\eqalign{
\b_2 &=-{1 \over 3}\left(3S_{d-1}\int_{{\cal R}_{12}} d^d x_2 r^{d-1} dr \<\cO(0)\cO(x_2)\cO(r e_1)\cO(\oo)\>\right)_{\de^{-1}}\cr
&= -{1 \over 3}\left(3S_{d-1} \int {dr \over r} r^{2d-2\Delta_\cO} \int_{{\cal R}} d^d x \<\cO(0)\cO(x)\cO(e_1)\cO(\oo)\>\right)_{\de^{-1}}\cr
&= -{1\over 6} \left(3S_{d-1} \int_{{\cal R}} d^d x \<\cO(0)\cO(x)\cO(e_1)\cO(\oo)\>\right)_{\de^{0}}\cr
}}
where ${\cal R}=\{x : |x|<1, |x|<|e_1-x|\}$.  In the second line, we have made the replacement $x_2\to r x$ and used a Ward identity for a four-point function under rescaling.  In the third line, we performed the integral over $r$ to obtain an overall pole in $\de$.\foot{The conformal cross-ratio $z$ in this configuration is given by $z=(x\cdot e_1)+i|x-(x\cdot e_1)e_1|$, so that ${\cal R}=\{z:|z|<1,|z|<|1-z|\}$.}

The remaining integral has a single divergence as $x\to 0$. (Other divergences are not present because $x$ is restricted to the region ${\cal R}$, away from the other operator insertions.)  The form of this divergence can be computed from the $\cO$ term in the $\cO\times\cO$ OPE, and it can be subtracted to give
\eqn\betatwoms{
\b_2 = -{1 \over 6} 3S_{d-1} \left(\int_{\cal R} d^d x \<\cO(0)\cO(x)\cO(e_1)\cO(\oo)\> - {S_{d-1} C_{\cO\cO\cO}^2 \over \de}\right).
}
Since in~\msbetacoefficients\ we are instructed to extract the term $\delta^{-1}$, in~\betatwoms\ the subtraction is again with the OPE coefficient evaluated at $\delta=0$. (As we have explained in the previous subsection, other prescriptions would differ by contact terms in the OPE of $\cO\CO$ but those cancel out in the final answer at this order.) If additional relevant operators are present in the $\cO\times\cO$ OPE, we isolate their contributions and integrate by analytic continuation in their dimensions. This is equivalent to cancelling the associated power-law divergences with counterterms. Although it is not obvious, one can check by isolating the contribution of $\cO$ in the $\cO\times \cO$ OPE that \betatwoms\ agrees with \betatwo\ and \betatwoimp\ in the limit $\de\to 0$.

The form \betatwoms\ is well-suited for computation using the conformal block expansion.  Because the integral is restricted to the region ${\cal R}$, we only need to keep a few terms in the expansion of $\<\cO\cO\cO\cO\>$ to get a good approximation. 

\subsec{Anomalous Dimensions}

It remains to review operator renormalization and anomalous dimensions in conformal perturbation theory.
Consider some operator $\Phi$ and let us deform the action by $\lambda\int d^dx\,\Phi(x)$ (in addition to the deformation by $\cO$). As before, we compute the overlap $\<\Phi|0\>_{\lambda,g,V}$ and demand that the result is independent of the UV cutoff. We keep terms that are linear in $\lambda$ but all orders in $g$. Let us assume $\Phi$ is canonically normalized and has the OPE
\eqn\OPEmixed{\cO(x)\Phi(0) \sim {C_{\Phi\Phi \CO}\over x^{\Delta_\cO}}\Phi(0)+\dots.}
We discuss here the marginal case $\Delta_\cO=d$.

We thus have, analogously to~\freeenexp,
\eqn\lambdaexp{\eqalign{\<\Phi|0\>_{\lambda,g,V}&=V\lambda \langle \Phi(0)\Phi(\infty))\rangle +V\lambda g \int d^dx \langle\Phi(0) \CO(x)\Phi(\infty)\rangle\cr
&~~ +\half V \lambda g^2 \int d^dx_1d^dx_2\langle\Phi(0)\CO(x_1)\CO(x_2)\Phi(\infty)\rangle +\dots.}}
Performing the same change of variables as before gives
\eqn\lambdaexp{\eqalign{
\<\Phi|0\>_{\lambda,g,V} &= V\lambda +V\lambda g \int d^dx \langle\Phi(0) \CO(x)\Phi(\infty)\rangle \cr
&~~+\half V \lambda g^2 S_{d-1}\int {d\rho\over \rho}\int d^dx\langle\Phi(0)\CO(e_1)\CO(x)\Phi(\infty)\rangle +\dots.}}
The log divergences in the four-point function are given by
\eqn\integratedim{\eqalign{
&S_{d-1}\int {d\rho\over \rho}\int d^dx\langle\Phi(0)\CO(e_1)\CO(x)\Phi(\infty)\rangle\cr
&= S^2_{d-1}\left(C^2_{\Phi\Phi \CO}+\half C_{\CO\CO\CO}C_{\Phi\Phi \CO}\right)\log^2(\mu)+A'\log(\mu).
}}
This is consistent with the beta function 
\eqn\consistentwithbetafn{\beta_\lambda=-\lambda \left(S_{d-1}C_{\Phi\Phi \CO}g+\half A' g^2+\dots\right),} and with $\beta_g$ as in~\betatwo. The anomalous dimension of $\Phi$ is  given by taking a derivative of $\beta_\lambda$, giving 
\eqn\Udim{\eqalign{
\Delta_\Phi(g) &= \Delta_\Phi + \gamma_{\Phi,1}g+\gamma_{\Phi,2}g^2 + \dots, \cr
\gamma_{\Phi,1} &= -S_{d-1} C_{\Phi\Phi\cO},\cr
\gamma_{\Phi,2} &= -{1 \over 2} A'.
}}

Unlike the beta function to second order in conformal perturbation theory, $\gamma_{\Phi,2}$ is scheme dependent. Indeed, consider the contact term $\Phi(x)\cO(0)\sim \alpha\delta^{(d)}(x)\Phi(x)$. From~\lambdaexp\ we readily see that this corresponds to the change of variables $\lambda\rightarrow \lambda+\alpha\lambda g+...$. As long as $\gamma_{\Phi,1}\neq 0$ (i.e. $C_{\Phi\Phi\cO}\neq 0$), this renders $\gamma_{\Phi,2}$ scheme dependent; the contact term $\Phi(x)\cO(0)\sim \alpha\delta^{(d)}(x)\Phi(x)$ shifts it as $\gamma_{\Phi,2}\rightarrow\gamma_{\Phi,2}+\alpha \gamma_{\Phi,1}$.

Therefore, we must choose some scheme when providing a formula for $\gamma_{\Phi,2}$. One possible choice is to cut a small hole around $\Phi(0)$ or, equivalently, choose $\alpha=0$. 
As before we can replace $\<\Phi\cO\cO\Phi\>$ with the connected correlator and subtract a contribution from generalized free field theory deformed by a double trace operator. The procedure is explained in appendix A.
One finds 
\eqn\finalA{\eqalign{
\gamma_{\Phi,2}&=-{S_{d-1} \over 2}\int d^dx \left[\langle\Phi(0)\CO(e_1)\CO(x)\Phi(\infty)\rangle_c
- {\left(C_{\Phi\Phi \CO}^2-\half  C_{\Phi \Phi \CO}C_{\CO\CO\CO}\right)\over x^d}\right.\cr
& \left.\qquad\qquad\qquad\qquad\qquad- {1 \over 2}{ C_{\Phi \Phi \CO}C_{\CO\CO\CO}  \over (x-e_1)^d}
 - {1 \over 2} {C_{\Phi \Phi \CO}C_{\CO\CO\CO}\over x^d(x-e_1)^d}\right].
 }}
If there are additional singular terms in the various OPE channels above, they are subtracted just like in~\betatwoimp.

In the minimal subtraction scheme with $\Delta_\cO=d-\de$, we have
\eqn\gammams{\eqalign{
\gamma_{\Phi,2} &= -{S_{d-1} \over 2}
\left[
2\left(\int_{\cal R} d^d x\<\Phi(0)\cO(x)\cO(e_1)\Phi(\oo)\>-{S_{d-1} C_{\Phi\Phi\cO}^2 \over \de}\right)\right.\cr
&\qquad\qquad~~~+\left.\left(\int_{\cal R} d^d x \<\cO(0)\cO(x)\Phi(e_1)\Phi(\oo)\> - {S_{d-1} C_{\Phi\Phi\cO}C_{\cO\cO\cO} \over \de}\right)\right].
}}

\subsec{Finite RG Flows and Scheme Dependence}

Finally, let us comment on which quantities are physically meaningful when $\de$ is small but fixed.  In this case, conformal perturbation theory is only a controlled approximation at distances $x\ll 1/\mu$, where $\mu$ is the scale associated with the deformation.  In particular, it does not give a controlled expansion for IR quantities like scaling dimensions in the deformed theory.  However, we can proceed (non-rigorously) as follows.  Let us pretend that a continuous family of CFTs with $\de\to 0$ exists, but that we only know the CFT data at one (small) value $\de=\de_*$.  The data at $\de=0$ (where conformal perturbation theory is well-defined) could then differ by $O(\de_*)$.  In practice, this means we can trust quantities computed with minimal subtraction at leading nonzero order. At subleading order, there are ambiguities due to our ignorance. 

In general, only the leading $\b$-function coefficient $\beta_1$ and leading anomalous dimension $\gamma_{\Phi,1}$ are unambiguous in this case.  This can be seen explicitly in \gammams: if we redefine $C_{\cO\cO\cO}$ and $C_{\Phi\Phi\cO}$ at order $\de$, this gives a nontrivial contribution to $\gamma_{\Phi,2}$ at order $\de^0$.  However, if $C_{\Phi\Phi\cO}$ vanishes, then $\gamma_{\Phi,2}$ becomes unambiguous.  We will encounter both situations below.

\newsec{The Random-Bond Ising Model with $d=4-\half \tilde\varepsilon$}
\seclab\randombondisinginfourminuseps

This section begins by studying the random-bond Ising model  in $d=4$~\dfour. To that end, we use conformal perturbation theory at one-loop, as reviewed in the previous section. We then discuss the model in $d=4-\half \tilde\varepsilon$ dimensions.

All we need for the present study are the OPE coefficients in free field theory between various products of the operator $\epsilon={1\over \sqrt 2} \phi^2$ and $\epsilon^2=\half \phi^4$. Here is a summary of all the  OPE coefficients pertinent to our cosiderations (some terms are omitted since they will not be important)
\eqn\allOPE{\eqalign{& \epsilon(x)\epsilon(0)\sim {1\over x^4}+{2\sqrt 2\over x^2}\epsilon(0)+\epsilon^2(0),\cr &  \epsilon(x)\epsilon^2(0)\sim  {6\over x^4} \epsilon (0),\cr & 
\epsilon^2(x)\epsilon^2(0)\sim  {6\over x^8}+ {36\over x^4}\epsilon^2(0). }}

The interaction term in~\dfour\ can be written as \eqn\interactionwritten{\int d^4x\left[\left(-\lambda^2+\half c^2\right)\sum_A\epsilon^2_A+\half c^2\sum_{A\neq B}\epsilon_A\epsilon_B\right].}
Using the OPE~\allOPE\ and implementing conformal perturbation theory at one loop as explained in the previous section we find that 
\eqn\Betas{\eqalign{& {d\over d\log\mu} \left(-\lambda^2+\half c^2\right)=-18 \left(-\lambda^2+\half c^2\right)^2 -\half (n-1) c^4 
,\cr &
\half{d\over d\log\mu}  c^2 = -6\left(-\lambda^2+\half c^2\right)c^2 -\half  (n+2) c^4. } }
Observe that in terms of the original couplings we have  
\eqn\Betasi{\eqalign{& {d\over d\log\mu} \lambda^2=6\lambda^2\left(3 \lambda^2-2c^2 \right)
,\cr &
{d\over d\log\mu}  c^2 = c^2\left(12\lambda^2-(n+8) c^2\right). } }
The reason that the coupling $c^2$ does not by itself lead to a term proportional to $\lambda$ and vice versa is easily understood from symmetries: If $c^2=0$ then it must remain so because the $n$ copies are decoupled. If $\lambda^2=0$ it must remain so because the coupling $c^2$ respects an $O(n)$ symmetry (and not just a permutation symmetry) which is clearly visible in terms of the microscopic fields. 

Let us now take $n\rightarrow 0$ as appropriate for the disordered theory. We get 
\eqn\Betadis{\eqalign{& {d\over d\log\mu} \lambda^2=6\lambda^2\left(3 \lambda^2-2c^2 \right)
,\cr &
{d\over d\log\mu}  c^2 = 4 c^2\left(3\lambda^2-2 c^2\right). } }

\item{1.} If disorder is absent, then the coupling $\lambda^2$ is marginally irrelevant, which, of course, is very well known.

\item{2.} If we take pure Mean Field Theory  (i.e.\ $\lambda^2=0$) in $d=4$ and introduce disorder in temperature, the theory, at least initially, flows to strong coupling because the beta function for $c^2$ is negative!

\item{3.} Disorder renders the coupling $\lambda^2$ relevant, as the dimension of the interaction proportional to $\lambda^2$ becomes $\Delta_{\epsilon^2}=4-12c^2$.

\item{4.} The one-loop approximation has a line of fixed points with $2c^2=3\lambda^2$. This degeneracy disappears at higher orders. This line of fixed points is attractive for $\lambda^2$ but unstable under disorder perturbations.

\medskip

We thus conclude that,  strictly speaking, the random-bond mean-field-theory in $d=4$ is {\it non-perturbative} in the infrared. The fixed points on the line $3\lambda^2=2c^2$ are unstable.  Furthermore, they disappear at higher order.

Let us now study the system in $d=4-\half \tilde \varepsilon$ dimensions. For that, to leading order, we only need to add a tree-level term to the beta functions: 
\eqn\Betadis{\eqalign{& {d\over d\log\mu} \lambda^2=-\tilde\varepsilon \lambda^2+6\lambda^2\left(3 \lambda^2-2c^2 \right)
,\cr &
{d\over d\log\mu}  c^2 = -\tilde\varepsilon c^2+4 c^2\left(3\lambda^2-2 c^2\right). } }
 Unfortunately, because the one-loop approximation has a line of fixed points, there is no solution with $\lambda^2,c^2\sim\tilde\varepsilon$.

 If one assumes $\lambda^2=0$ one finds that the theory flows to strong coupling. If one assumes $c^2=0$ then one finds the Wilson-Fisher fixed point with $\lambda^2={1\over 18}\tilde\varepsilon$. At this Wilson-Fisher fixed point the anomalous dimension of disorder is $4-{1\over 3}\tilde \varepsilon$, which means that the Wilson-Fisher fixed point is now a UV fixed point. This is consistent with our general picture that disorder is relevant in the Ising model everywhere between $2<d< 4$ (i.e.\ that the heat capacity in the pure Ising model is positive for all $2<d< 4$). See Fig~1.

A weakly coupled infrared fixed point can be found when one includes higher-order corrections to~\Betadis, but this results in an expansion in $\sqrt {\tilde\varepsilon}$, which is less effective than the standard $\tilde \epsilon$ expansion. For example,  the first few nontrivial corrections to the infrared heat capacity in $d=3$ do not give the correct sign. See~\fiveloops\ and references therein.

\newsec{Disordered Theories and Conformal Perturbation Theory}
\seclab\generalresultsfromcpt

In this section we apply the results of sections~\thereplicamethod\ and \reviewofconformalperturbationtheory\ to a general conformal field theory in which disorder is a marginal or a nearly-marginal perturbation. The results we obtain here could be useful in a variety of models, and in particular, they will be useful in the two sections that follow.

To set up the problem, recall that we are considering the following partition function 
\eqn\dtwosecfive{W_n=\int\prod_{A=1}^n [{\cal D}\sigma_A]e^{-\sum_AS_A+{c^2\over 2}\int d^dx \sum_{A\neq B}\epsilon_A(x)\epsilon_B(x)},}
with $S_A$ being the action of some $d$-dimensional conformal field theory in which there is an operator, $\epsilon_A$, of dimension close to $d/2$. The $n$-copies are identical and decoupled if $c^2=0$.

We assume that the unperturbed pure CFT does not have an operator $\epsilon^2$, i.e.\ we assume that the OPE in the pure theory is of the form 
\eqn\OPEenergy{\epsilon(x)\epsilon(0)={1\over x^{2\Delta_\e}}+ {C_{\e\e\e} \over x^{\Delta_\e}}\epsilon(0)+{\rm vanishing\ as\ }x\to 0.}
We assume that $\Delta_\epsilon$ is either $d/2$ or parametrically close to $d/2$. 

The assumption~\OPEenergy\ renders our discussion inapplicable for non-generic theories like that of section~\randombondisinginfourminuseps\ (which is mean-field theory, so the operator $\epsilon^2$ exists) or large-$N$ theories, where if $\epsilon$ is a single trace operator then $\epsilon^2$ is a double-trace operator. But our discussion is relevant in many other cases. The non-generic cases can be treated separately as in section~\randombondisinginfourminuseps. The case of $\epsilon$ being a single-trace operator in a large-$N$ theory was treated in~\AharonyAEA. See also references therein.

For what follows we will need the various OPEs in which the interaction term $\sum_{A\neq B}\epsilon_A(x)\epsilon_B(x)$ participates. We find
\eqn\OPEsnco{
\eqalign{
\sum_{A\neq B}\epsilon_A\epsilon_B\sum_{C\neq D}\epsilon_C\epsilon_D &\sim 
{2n(n-1)\over x^{4\Delta_\epsilon}}
+
{4C_{\e\e\e}(n-1) \over x^{3\Delta_\epsilon}}\sum_A\epsilon_A+{4(n-2)+2C_{\e\e\e}^2 \over x^{2\Delta_\epsilon}}\sum_{A\neq B}\epsilon_A\epsilon_B+\dots,\cr
\sum_A \epsilon_A\sum_{B\neq C}\epsilon_B\epsilon_C
&\sim {2(n-1)\over x^{2\Delta_\epsilon}}\sum_A\epsilon_A +{2C_{\e\e\e}\over x^{\Delta_\epsilon}}\sum_{A\neq B} \epsilon_A\epsilon_B+\dots,\cr 
\sum_A \epsilon_A\sum_B \epsilon_B
&\sim {n\over x^{2\Delta_\epsilon}}+\dots.}}
Following section 3, we can define a canonically normalized operator $\cO(x)$, and associated coupling constant $g$,
\eqn\defofO{\eqalign{
\cO(x) &\equiv \frac{1}{\sqrt{2n(n-1)}}\sum_{A\neq B} \e_A(x)\e_B(x),\cr
g &= \frac{c^2}{2} \sqrt{2n(n-1)}.
}}
It is also useful to define a canonically normalized energy operator in the replica theory
\eqn\O{\eqalign{ \cE={1\over \sqrt n}    \sum_A\epsilon_A.  } }
We can read off the OPE coefficients in this basis from \OPEsnco:
\eqn\normalizedOPE{\eqalign{& C_{\cO\cO\cO}={4(n-2)+2C_{\e\e\e}^2\over\sqrt {2n(n-1)}  } , \cr&
C_{\cE\cE\cO}={\sqrt {2(n-1)} \over   \sqrt {n } }, \cr& 
C_{\cE\cO\cO}={2C_{\e\e\e}\over \sqrt n}.
}}

We also introduce the canonically-normalized spin operator in the replica theory
\eqn\Spin{\cS(x)=\frac 1 {\sqrt n}\sum_A \sigma_A(x).}
We assume that in the pure theory,
\eqn\OPEspinpure{\sigma(x)\sigma(0)\sim {1\over x^{2\Delta_\sigma}}+ {C_{\sigma\sigma\epsilon}\over x^{2\Delta_\sigma-\Delta_\epsilon}}\epsilon(0)+\dots.}
It is then useful to note that 
\eqn\normalizedOPEi{\eqalign{& C_{\cS\cS\cO}=0,\cr &
C_{\cS\cO\cO}=0.
}}
The latter follows immediately from the $Z_2$ symmetry in the pure model.

We are now ready to proceed with one- and two-loop computations in the replicated theory.

\subsec{One Loop}

First, let us assume $\Delta_\epsilon=d/2$. Then, using~\betaone\ and the OPE coefficient $C_{\cO\cO\cO}$ found in~\normalizedOPE\ we obtain the beta function
\eqn\betafun{{d\over d\log(\mu)}\left(\frac{c^2}{2}\right)=-\half S_{d-1}\left(4(n-2)+2C_{\e\e\e}^2\right)\left(\frac{c^2}{2}\right)^2+O(c^6).}
The beta function for the physical disordered theory is obtained by simply substituting $n=0$ in~\betafun\ 
\eqn\betafunneqz{ {d\over d\log(\mu)}\left({1\over c^2}\right)=-S_{d-1}\left(2-\half C_{\e\e\e}^2\right)+O(c^2).} 
This shows that in the marginal case, unless the OPE coefficient $C_{\e\e\e}$ is larger than 2, disorder decays back to the pure CFT. It is therefore {\it marginally irrelevant} as long as \eqn\marg{|C_{\e\e\e}|<2.}
It would be nice to know if there are CFTs in which this bound is violated. This should be possible to answer with numerical bootstrap techniques. In the 2d Ising model, $C_{\e\e\e}=0$ as a result of the duality between high and low temperatures. Therefore, disorder is marginally irrelevant. We will discuss this in much more detail soon. For some early work on the disordered $d=2$ model, see~\VV.

Let us now denote
\eqn\defofdelta{
\delta \equiv d-2\Delta_\epsilon
}
and imagine that $\delta$ is parametrically small (i.e.\ the heat capacity is parametrically small). As in \msbetafunctiontwo, we add to the beta function a linear piece in $c^2$:
\eqn\betagain{{dc^2\over d\log(\mu)} = -\delta c^2+S_{d-1}\left(2-{1\over 2}C_{\e\e\e}^2\right)c^4+O(c^6).}
As long as~\marg\ holds and $0<\delta \ll 1$, this leads to a weakly-coupled, disordered, fixed point at long distances  with
\eqn\fixeddis{S_{d-1}c_*^2\simeq {\delta \over 2-\half C_{\e\e\e}^2}.}
Indeed, if $\delta \ll1$, the higher order corrections are suppressed.

We now calculate the heat-capacity exponent of this infrared disordered theory. This is done by calculating the infrared dimension of $\epsilon$  in the disordered theory. In the replica trick, this corresponds to calculating the infrared dimension of $\cE(x)=\frac 1 {\sqrt n}\sum_A \epsilon_A(x)$.
Using \Udim,
 the dimension in the deformed theory is given by
\eqn\deformeddimensionofE{\eqalign{
\Delta_{\cE}^{\rm IR} &= \Delta_\e - S_{d-1} C_{\cE\cE\cO} g_* + \dots\cr
&= \Delta_\e - (n-1)S_{d-1}c_*^2 + \dots\cr
\Delta_\cE^{\rm IR}(n=0)&= \Delta_\e + \frac{\de}{2-\half C_{\e\e\e}^2}+\dots,
}}
where in the last line we set $n=0$ and used \fixeddis.
We immediately see that, to this order, the dimension of the energy operator in the infrared is always bigger than in the ultraviolet.

The disordered heat-capacity exponent in the infrared (at leading order in $\delta$) is
\eqn\irheatca{\alpha^{{\rm IR}}=-{C_{\e\e\e}^2 \delta\over {2d-(d-\Delta_\epsilon)   C_{\e\e\e}^2 }}+\dots.}
Since this formula is only to be trusted at leading order in $\delta=d-2\Delta_\epsilon$, we can simplify the denominator and get
\eqn\irheatcai{\alpha^{{\rm IR}}=-{C_{\e\e\e}^2 \delta \over {d(2-\half C_{\e\e\e}^2) }}+\dots.}
Note an interesting fact: As long as the OPE coefficient $C_{\e\e\e}$ satisfies~\marg, the infrared heat capacity exponent satisfies $\alpha^{{\rm IR}}\leq 0$. As reviewed in the introduction, this property of disordered fixed points is expected on general grounds. 

Let us also consider the magnetic susceptibility exponent $\gamma^{{\rm IR}}$. This is determined by the dimension of the spin operator $\sigma(x)$ in the disordered theory, or correspondingly the operator $\cS(x)=\frac 1 {\sqrt n}\sum_A \sigma_A(x)$ in the replica theory.   
Since the three-point coefficient $C_{\cS\cS\cO}$ vanishes~\normalizedOPEi, the one-loop correction to the anomalous dimension of the spin operator vanishes as well.  

Recalling that $\gamma={d-2\Delta_\sigma\over d-\Delta_\epsilon}$, 
we therefore find that to the leading order in $\delta$ 
\eqn\gammacrit{{\gamma^{{\rm IR}}\over \gamma^{UV} }=1+{1\over d}{\delta \over (1-{1\over 4} C_{\e\e\e}^2)}+\dots}
An interesting consequence of these results is that as long as~\marg\ is upheld, we have that $\gamma^{{\rm IR}}>\gamma^{UV}$ to leading order. As we will mention again in the section that follows, this is consistent with experimental results. 

\subsec{Two Loops}

For a two-loop computation in the replica theory we need to consider the following integral 
\eqn\replicaeq{\eqalign{&\int d^dx \biggl[\left\langle\sum_{A\neq B} \epsilon_A\epsilon_B(0)\sum_{C\neq D}\epsilon_C\epsilon_D(e_1)\sum_{E\neq F}\epsilon_E\epsilon_F(x)\sum_{G\neq H}\epsilon_G\epsilon_H(\infty)\right\rangle_c\cr&-n(n-1)(4(n-2)+2C_{\e\e\e}^2)^2\left({1\over x^d (x-e_1)^d} +{1\over x^d}+{1\over (x-e_1)^d}\right) \biggr],}}
where $e_1$ is a unit vector in the $1$ direction.
As instructed in~\betatwo, we have already subtracted the divergence coming from the OPE coefficient $C_{\cO\cO\cO}$. This integral is still divergent because of the OPE 
coefficient $C_{\cO\cO\cE}$ but this can be subtracted as well according to the prescription of footnote 5 which we utilize below. 

The four-point function in~\replicaeq\ can be expressed in terms of the  connected four-point functions in the Ising model
\eqn\Isingfour{\langle \epsilon(0) \epsilon(e_1)\epsilon(x)\epsilon_\infty\rangle_c =F(x).}
This computation is a little bit involved and we collect the main steps in appendix~B. The result after 
the dust settles is the following beta function of the replicated theory 
\eqn\betafunctionreplicated{\beta_{c^2}=-\left((n-2)+\half C_{\e\e\e}^2\right) c^4-{1\over 24}\CI(n)\  c^6+\dots, }
 with~$\CI(n)$ given by 
\eqn\replicaeqii{\eqalign{
\CI(n)=&\int d^dx \biggl[4F^2(x)
+8(n-1)F(x)\left(1+{1 \over x^{d}}+{1 \over (x-e_1)^{d}}\right)\cr&
\qquad\quad+16(n-2){C_{\e\e\e}^2\over (x-e_1)^{d/2}x^{d/2}}\left({1 \over (x-e_1)^{d/2}}+{1 \over x^{d/2}}+1\right)
 \cr&
\qquad\quad+8\left(n-1- C_{\e\e\e}^2(n-2)-{1\over 4}C_{\e\e\e}^4\right)\left({1 \over x^{d}}+{1 \over (x-e_1)^{d}}+{1 \over x^{d}(x-e_1)^{d}}\right)
\cr&
\qquad\quad -8(n-1)C_{\e\e\e}^2\left({1 \over x^{3d/2}}+{1 \over (x-e_1)^{3d/2}}+x^{d/2}\right)
 \biggr].}}
In the last line, we have applied the prescription in~\betatwoimp\ to subtract off the $x^{-3d/2}$ singularity originating from the OPE coefficient $C_{\cO\cO\cE}$.
 
Since we are ultimately interested in the disordered $n=0$ theory, we simply plug in $n=0$ in~\replicaeqii\ and we conclude 
\eqn\finaltwoloopbeta{\beta_{c^2}^{n=0}=\left(2-\half C_{\e\e\e}^2\right) c^4-{1\over 24} \CI(0) c^6+\dots.}

It is useful to consider also the renormalization of the dimensions of $\cE$ and $\cS$ at two loops. This is done by considering the integrated correlation functions 
\eqn\replicaeqi{\eqalign{
&\int d^d x \<\cE(0)\cO(e_1)\cO(x)\cE(\oo)\>_c,\cr
&\int d^d x \<\cS(0)\cO(e_1)\cO(x)\cS(\oo)\>_c,
}}
with proper subtractions. These correlation functions can be reduced to two-, three-, and four-point functions in the pure Ising model. The details are in appendix~B. One finds the following formulae for the dimensions of $\epsilon$ and $\sigma$:
\eqn\sigmaepstwoloops{\eqalign{
\Delta_\cE^{\rm IR}&=\Delta_\epsilon-(n-1)S_{d-1} c^2+{\cal B} c^4,\cr
\Delta_\cS^{\rm IR}&=\Delta_\sigma+{\cal D} c^4,
}}
where 
\eqn\epsilontwoloops{\eqalign{{\cal B}=-S_{d-1}&(n-1)\int d^dx
\biggl[{F(x)\over (x-e_1)^d}+{C_{\e\e\e}^2\over x^{d/2}(x-e_1)^d}
\cr
&+(n-1)\left({1\over (x-e_1)^d}+{1\over x^d (x-e_1)^d}\right)
-{C_{\e\e\e}^2} {1\over (x-e_1)^{3d/2} }
\cr
& -{1\over 2x^d}\left(1-\half C_{\e\e\e}^2\right) - {1\over 2}\left(n-2+\half C_{\e\e\e}^2\right)\left({1\over (x-e_1)^d} +{1\over x^d(x-e_1)^d}\right)
 \biggr],}}
and
\eqn\sigmatwoloopsi{{\cal D}=-S_{d-1}(n-1)\int d^dx \left[ {G(x)\over (x-e_1)^d}-C_{\e\e\e} C_{\sigma \sigma \epsilon} {1\over (x-e_1)^{3d/2}}\right].}
Above we defined $G(x)=\langle \sigma(0)\epsilon(1)\epsilon(x)\sigma_\infty\rangle_c $. The integrals defining $\CB$ and $\CD$ converge.
One can plug $n=0$ and obtain slightly simplified formulae from which the two-loop contributions to the dimensions of the energy and spin operators in the disordered theory can be obtained.

\newsec{The Random-Bond Ising Model with $d=2+\varepsilon$}
\seclab\randombondintwopluseps

We begin with the observation that for small $\varepsilon$ the pure and disordered fixed points are parametrically close. Indeed, using~\ElShowkNIA\ we can compute the dimension of the energy operator in the pure theory to first order 
\eqn\dimenergy{\Delta_\epsilon=1+0.3\varepsilon+\CO(\varepsilon^2).}
And therefore the heat capacity exponent 
\eqn\twoplusepsheat{\alpha\equiv {D-2\Delta_\epsilon\over D-\Delta_\epsilon}\sim 0.4\varepsilon+\CO(\varepsilon^2).}
This is positive for $\varepsilon>0$ and therefore disorder is slightly relevant for small, positive, $\varepsilon$.
We can thus try to compute various properties of the disordered theory in the infrared. 
Because of the general result~\irheatca, we find that the infrared heat capacity exponent vanishes to leading order in $\varepsilon$. This is because $C_{\e\e\e}=0$ in $d=2$ due to the high-low temperature duality in the Ising model. Therefore, we get for the disordered heat capacity exponent
\eqn\disorderedheatcapacity{\alpha^{\rm IR}(\varepsilon) \sim \varepsilon^2.}
In other words, the leading order in $\varepsilon$ prediction for the heat capacity exponent in the disordered theory is that the heat capacity exponent vanishes. 
This is similar to the degeneracy of the perturbation in $q-2$ in~\LudwigRK. This degeneracy in the first order in $\varepsilon$ has already appeared in section 8 of~\Cardy.

It would be nice to compute the coefficient of $\varepsilon^2$. Clearly the sign of this contribution is crucial. For that we need to consider first the disordered beta function in $d=2$ to two loops. Using section 3, we find a result in agreement with~\LudwigRK:
\eqn\betaIsing{{dc^2\over d\log\mu}=-2\pi(n-2)c^4+(n-2)(2\pi)^2c^6+\dots}  
\eqn\betaIsingi{{d\lambda \over d\log\mu}=-\lambda\left( 1+2\pi (n-1)c^2-\half (n-1) (2\pi)^2 c^4\right)+\dots}  
where $\lambda$ corresponds to the coupling that generates correlation functions of the energy operator.

The one-loop terms are precisely consistent with our~\betafun\ and~\betagain. The vanishing of the beta function for $n=2$ is due to the fact that the deformation $\epsilon_1\epsilon_2$ is exactly marginal in this case (it is a $c=1$ model, called the Ashkin-Teller model).

Now let us imagine studying the theory at $d=2+\varepsilon$ with very small $\varepsilon$. We have proven above that to first order in $\varepsilon$ the heat capacity in the disordered fixed point vanishes.  To go to 2nd order in $\varepsilon$ we need to know the corrections of order $\varepsilon$ to the one-loop beta functions and we need to know the two-loop beta functions. For the two-loop beta functions we can use the $d=2$ Ising values. The error in doing that is of the order of $\varepsilon^3$. 

So it remains to discuss the corrections in $\varepsilon$ to the one-loop beta function. The one loop coefficient in~\betaIsing\ has in general a term $C_{\e\e\e}^2$, see~\betafun. $C_{\e\e\e}=0$ in the 2d Ising model. If we assume that $C_{\e\e\e}$ starts at order  $\varepsilon$ then we can therefore neglect this in our computation since this correction is formally of the order of a three-loop contribution. Hence, our beta functions take the form 

\eqn\betaIsing{{dc^2\over d\log\mu}=-(d-2\Delta_\epsilon)c^2-2\pi(n-2)c^4+(n-2)(2\pi)^2c^6+\dots}  
\eqn\betaIsingi{{d\lambda\over d\log\mu}=-\lambda\left( (d-\Delta_\epsilon)+2\pi (n-1)c^2-\half (n-1) (2\pi)^2 c^4\right)+\dots}  

We take $n\rightarrow 0$ and solve for $c^2$ at the fixed point 
\eqn\cstaratfixedpointtwod{4\pi c_*^2= d-2\Delta_\epsilon+\half (d-2\Delta_\epsilon)^2+\dots.}
And hence the infrared dimension of the energy operator is 
\eqn\infrareddimensionofenergyop{d-\Delta^{\rm IR}_\epsilon=d-\Delta_\epsilon-2\pi c_*^2+2\pi^2c_*^4+\dots.}
Plugging our solution for $c_*^2$ we find 
\eqn\plugginginsolnforcstar{\Delta_\epsilon^{\rm IR}-d/2={1\over 8} (d-2\Delta_\epsilon)^2+\dots.}
We can trust this formula to order $\varepsilon^2$ so we plug in the expression $d-2\Delta_\epsilon=2+\varepsilon-2-0.6\varepsilon=0.4\varepsilon$. Hence we find to order $\varepsilon^2$
\eqn\hencewefindtorderepstwo{\Delta_\epsilon^{\rm IR}=1+\half \varepsilon+0.02\varepsilon^2+\dots.}
The sign of the second order correction is correct (positive), and it leads to a negative heat capacity exponent which takes the following form in the expansion in $\varepsilon$: 
\eqn\alphairtwopluseps{\alpha^{\rm IR}={d-2\Delta_\epsilon\over d-\Delta_\epsilon}\sim - {0.04 }\varepsilon^2+\dots.}
Upon plugging $\varepsilon=1$ we get $\alpha^{\rm IR}=-0.04$, which is well within the range of experimental and Monte Carlo results as summarized in~\MC!

We finish by considering the dimension of the spin field $\sigma$ to the first nontrivial order in the epsilon expansion. As before, let us first consider the anomalous dimension of the spin field in $d=2$ as a function of the marginal disorder coupling, $c$. If we deform the action by the spin field with coefficient $M$, then we will find that the beta function of $M$ is given by  
\eqn\spin{{dM\over d\log\mu }=-M\left({15\over 8}+\CO(c^6)\right). }
The first nontrivial contribution arises only at third order in conformal perturbation theory! 
The vanishing of the first order in perturbation theory is obvious from the fact that $\sigma$ and $\epsilon$ are orthogonal operators. This has been already explained above~\gammacrit. 

The vanishing of the second order in conformal perturbation theory is less obvious. The term of order $c^4$ is given by $\CD$ in~\sigmatwoloopsi.  In the 2d Ising model, we have
\eqn\twodisingsigsigepseps{\eqalign{
G(z,\bar z) = \<\sigma(0)\epsilon(z)\epsilon(1)\sigma(\oo)\>_c &= \frac{|1+z|^2}{4|z||1-z|^2} - \frac{1}{|1-z|^2}.
}}
Plugging this into \sigmatwoloopsi, we have checked numerically that the resulting integral vanishes. It would be nice to understand why this happens.
%
Vanishing of the two-loop contribution also follows from the analysis of~\Dotsenko, which can be used to extract the $c^6$ term in the anomalous dimension of $\sigma$.

We now need to discuss how~\spin\ is modified in $2+\varepsilon$ dimensions. The first order in conformal perturbation theory vanishes for all $\varepsilon$ due to~\sigmaepstwoloops\ but the second order may not. Hence, the second order may in general take the form $\sim \varepsilon c^4$, which in our Wilson-Fisher-like fixed point would be comparable with $c^6$. Therefore, we cannot compute this correction without additional work, but we can already conclude that 
\eqn\thirdorder{\Delta_\sigma^{\rm IR}(\varepsilon)-\Delta_\sigma(\varepsilon)=\CO(\varepsilon^3).}

Thus we can use~\thirdorder\ to give the expansion in $\varepsilon$ for $\gamma^{\rm IR}/\gamma^{\rm UV}$ containing the first two nontrivial orders in $\varepsilon$. Indeed, 
\eqn\indeed{{\gamma^{\rm IR}\over \gamma^{\rm UV}}={(d-2\Delta^{\rm IR}_\sigma)(d-\Delta_\epsilon)\over (d-\Delta^{\rm IR}_\epsilon)(d-2\Delta_\sigma)}.}
Therefore, if we neglect corrections of order $\varepsilon^3$, we find: 
\eqn\orderthree{{\gamma^{\rm IR}\over \gamma^{\rm UV}}={d-\Delta_\epsilon\over d-\Delta^{\rm IR}_\epsilon}.}
This is generally a positive number because $\Delta_\epsilon^{\rm IR}$ increases in the infrared relative to $\Delta_\epsilon$.

If we just use~\orderthree\ directly in $d=3$ we find that the ratio is $\gamma^{\rm IR}/\gamma^{\rm UV} = {1.59/1.48}\sim 1.075-1.08$
while the experimentally measured ratio is around $1.09$, so the agreement with experiment is rather good.

 \newsec{An Uncontrolled Computation in $d=3$}
\seclab\uncontrolledthreed

In the 3d Ising model, the energy operator has dimension $\Delta_{\epsilon}\sim 1.41$, which is rather close to the Harris bound $d/2=1.5$. One could therefore attempt to compute observables in a perturbative expansion in $\delta=d-2\Delta_\e\approx 0.18$.  Note that in the 3d Ising model, there is no nearly marginal operator in the OPE of the energy operator with itself. Indeed, the next $\Z_2$-even operator has dimension $\sim 3.84$ \ElShowkHT. (We give a more precise estimate of this dimension below.)  Therefore, the formalism of the previous section applies.

Of course a perturbative expansion in $0.18$ is not rigorously justified, but given the observations above, it might be a reasonable approximation. (In the next section we explain how this can be turned into a rigorous expansion.) The idea behind this approximation is summarized in Fig~2: Since $\delta$ in the Ising model is numerically small, and because there is no operator of the type $\epsilon^2(x)$, the disordered Ising fixed point may be viewed as being close in the RG sense to the pure nontrivial fixed point. 

\bigskip
\vbox{
\offinterlineskip
\lineskip=3pt
\epsfxsize=2.0in \centerline{\epsfbox{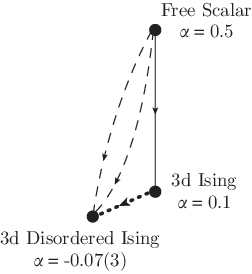}}
\nobreak\vskip0.4cm {
{\it \vbox{\noindent {\bf Figure 2:}
{\it A schematic picture of the renormalization group flow between the free scalar theory and the 3d pure and disordered Ising model, with disorder growing from right to left.}}}
}
}
\bigskip

It is also interesting that we can perform this purely 3d analysis at all.
To apply conformal perturbation theory to the 3d Ising model, we take dimensions and OPE coefficients from the conformal bootstrap.  These results will enable determinations of $\Delta_\e^{\rm IR}$ and $\Delta_\s^{\rm IR}$ (and the corresponding critical exponents) at one and two loops, respectively.

\subsec{One Loop}

To evaluate $\alpha^{\rm IR}$ using \irheatcai, we need $\Delta_\e$ and the OPE coefficient $C_{\epsilon\epsilon\epsilon}$ in the 3d Ising model. Using the techniques of~\refs{\KosBKA,\SimmonsDuffinQMA} we find\foot{In reference~\CaselleCSA\ the value $C_{\epsilon\epsilon\epsilon}=1.45\pm0.30$ is obtained. In~\CostagliolaIER\ the value $C_{\epsilon\epsilon\epsilon}= 1.32 \pm 0.15$ is reported.} \foot{Note that in the $d=2$ Ising model, $C_{\epsilon\epsilon\epsilon}=0$ by the Kramers-Wannier duality. In $d=4$, Mean Field Theory is valid and we get~\allOPE\ $C_{\epsilon\epsilon\epsilon}=2\sqrt2$. The average of these two values is in the right ballpark for $C_{\e\e\e}$ in $d=3$.}
\eqn\OPEIsing{\eqalign{
\Delta_\e &= 1.412625(10),\cr
C_{\epsilon\epsilon\epsilon}& =1.532435(19).
}
}
More detail on the computation of $\Delta_\e$ and $C_{\e\e\e}$ will be reported in \ONFuture. Substituting these values into \irheatcai\ gives
\eqn\oneloopalphaestimate{
\alpha^{\rm IR} \approx -0.166+O(\de^2).
}
Measurements of the heat capacity in the disordered Ising model yield results around $\alpha^{{\rm IR}}=-0.05$ (see~\MC\ and references therein) and $\alpha^{{\rm IR}}=-0.1$ (see~\Belanger\ and references therein). We do not quote errors on these numbers since there seems to be a significant spread of results in the literature.  The theoretical result \oneloopalphaestimate\ is somewhat larger in magnitude than the experimental results, though roughly of the same order (and with the correct sign).

It is interesting to turn the logic around and use the experimental measurements of disordered exponents as a way of estimating $C_{\e\e\e}$ in the pure theory.
The quoted values of $\alpha^{{\rm IR}}$ roughly correspond to
\eqn\quotedvalues{C_{\epsilon\epsilon\epsilon}\sim 1.2 - 1.35.}
This is a fairly satisfying result for a leading order approximation!  
Similarly, we can use~\gammacrit\ to obtain another estimate of $C_{\epsilon\epsilon\epsilon}$. The results quoted in the literature are around $\gamma^{{\rm IR}}=1.34$. In the pure theory we have $\gamma^{\rm UV}=1.23$. This leads to an estimate around $C_{\epsilon\epsilon\epsilon}=1.2$ which is again 
both satisfactory as a leading-order approximation and consistent with the estimate from the critical exponent $\alpha^{{\rm IR}}$.

It would be interesting to extend the analysis of this section also to correlators of the stress tensor. For comments about the stress tensor in disordered theories, see~\refs{\Gurarie,\StressCardy}. One can readily verify that the first correction due to disorder is at least of the order $\CO(c^4)$, i.e.\ it is quadratic in the heat-capacity exponent. This suggests the difference between the pure and disordered two-point function of the stress tensor will be numerically small in the 3d Ising model. 

\subsec{Two Loops}

The anomalous dimension of the spin operator 
\eqn\spinreminder{
\cS(x) = \frac{1}{\sqrt n}\sum_A \s_A(x).
}
vanishes at one loop, due to vanishing of the OPE coefficient $C_{\cS\cS\cO}$.  According to the discussion in subsection~3.4, the two-loop anomalous dimension $\gamma_{\cS,2}$ is then scheme-independent (up to three-loop corrections). We can compute it by considering the integrated four-point function $\<\cS\cO\cO\cS\>$ in the replicated theory.  This can in turn be written in terms of $\<\s\e\e\s\>$ in the pure theory,
\eqn\spinfourptreplica{\eqalign{
\<\cS(x_1)\cO(x_2)\cO(x_3)\cS(x_4)\> &=
\frac{1}{n}\frac{1}{x_{12}^{2 \Delta_\sigma } x_{34}^{2 \Delta_\sigma }x_{23}^{4 \Delta_\epsilon-2 \Delta_\sigma}}\cr
&
\qquad
 \times \left(2  u^{\frac{\Delta_\sigma -\Delta_\epsilon }{2}}v^{\Delta_\epsilon-\Delta_\sigma  } g_{\sigma \epsilon \epsilon \sigma }(u,v)
  +
  (n-2) \frac{u^{\Delta_\sigma}}{v^{\Delta_\sigma}} \right),
  }}
  where
\eqn\spinfourptpure{\eqalign{
\<\s(x_1)\e(x_2)\e(x_3)\s(x_4)\> &= \left(\frac{x_{24} x_{13}}{x_{14}^2}\right)^{\De_\s-\De_\e}\frac{g_{\s\e\e\s}(u,v)}{x_{12}^{\De_\e+\De_\s} x_{34}^{\De_\e+\De_\s}}.
}}

We will approximate the four-point function and $g_{\s\e\e\s}(u,v)$ using the conformal block expansion, truncated at some large value $\Delta_*$ of the dimension of the exchanged operator.  We have already written the integral \gammams\ in such a way that the conformal cross ratio $z$ is restricted to the region
\eqn\crossratioregion{
{\cal R} = \{z: |z|\leq 1~{\rm and}~|z|\leq |1-z|\}.
}
The error from truncating the conformal block expansion at $\Delta_*$ falls exponentially as $|\rho(z)|^{\Delta_*}$, where \refs{\PappadopuloJK,\RychkovLCA}
\eqn\definitionofrho{
\rho(z) \equiv \frac{z}{(1+\sqrt{1-z})^2}.
}
The error in the integral will fall asymptotically
\eqn\integralerror{
\max_{z\in{\cal R}} |\rho(z)|^{\Delta_*}=(2-\sqrt 3)^{\Delta_*}\approx 0.27^{\Delta_*},
}
where the maximum value of $|\rho(z)|$ is achieved at $z=\frac 1 2 \pm i\frac{\sqrt 3}2$.  In practice, the OPE coefficients in the Ising model are such that $\Delta_*=6$ already gives a good approximation for the integrals we need, to one part in $10^{-3}$.

Thus, to estimate our four-point functions in the region ${\cal R}$, we need the OPE coefficients and dimensions of low-lying operators.\foot{When the quantity to be integrated is a linear function of Ising four-point functions, then it is possible to obtain direct bounds on the integral without computing the operators that appear in the four-point function \LinWCG.  An advantage of this approach is that we obtain precise error bars on the result.  Here, we take the more straightforward route of approximating the entire function $g_{\s\e\e\s}(u,v)$ first and then integrating.}  The numerical bootstrap \RattazziPE\ has taught us that crossing symmetry and unitarity severely constrain these quantities.  The usual procedure is to study $N$ derivatives of the crossing equations around the point $z=\bar z = \half$.  These $N$ equations, together with unitarity, then determine $\Delta_{\s}$, $\Delta_{\e}$ and the OPE coefficients $C_{\s\s\e}$, $C_{\e\e\e}$ to high precision \refs{\ElShowkHT,\ElShowkDWA,\KosBKA,\SimmonsDuffinQMA,\ONFuture}.

We will need information about a few more operators in the conformal block expansion.
To set rigorous bounds on $k$ operator dimensions requires a $k$-dimensional scan, which quickly becomes infeasible as $k$ grows. So to estimate the low lying spectrum, we adopt the non-rigorous approach of \refs{\ElShowkHU,\ElShowkDWA}.  For each $N$, we compute a set of dimensions and OPE coefficients that solve $N$ derivatives of crossing equations.  As $N$ grows, some dimensions and OPE coefficients for low-lying operators begin to converge, giving an estimate for their values in the 3d Ising CFT.

To obtain a unique set of dimensions and OPE coefficients for each $N$, we can extremize some quantity \refs{\PolandWG,\ElShowkHU}. As we vary the parameters of the extremization problem (for example, the values of $\Delta_\sigma$ or $\Delta_\epsilon$, the quantity to be extremized, or the number of derivatives $N$), some operator dimensions remain stable, while others jump around. The ``stable" operators have relatively low dimension and large coefficients in the $\s\times\s$, $\s\times\e$ and $\e\times\e$ OPEs.  Presumably, the stable operators are actually present in the 3d Ising CFT, while the jumping operators come from the arbitrariness of the extremization procedure and disappear at the true Ising point, as discussed in \ElShowkDWA.  Fortunately for us, the operators with low dimension and large OPE coefficients are exactly the ones that are most important for our calculation.

We guess the ``stable" operators by computing a few example spectra and comparing them.  To compute a spectrum we set up a semidefinite program (SDP) using the techniques of \refs{\PolandEY,\KosTGA,\KosBKA}, and then solve it using {\tt SDPB} \SimmonsDuffinQMA. The spectrum and OPE coefficients can be extracted from a combination of the primal and dual solutions of the SDP. More detail on this process will be reported elsewhere. Our example spectra are computed by fixing $\Delta_\sigma,\Delta_\epsilon,$ and $\theta\equiv\arctan(C_{\e\e\e}/C_{\s\s\e})$ to some values in the allowed region \ONFuture\ and then maximizing $C_{\e\e\e}$. Specifically, we take 
\eqn\spectrumchoices{\eqalign{
(\De_\s,\De_\e,\theta)\in \{
&
(0.51814834, 1.4126195, 0.96925883), (0.51814861, 1.4126207, 0.96925745),
\cr
&
(0.51814880, 1.4126238, 0.96926045), (0.51814900, 1.4126253, 0.96926130),
\cr
&
(0.51814953, 1.4126321, 0.96926690), (0.51814957, 1.4126322, 0.96926711),
\cr
&
(0.51814983, 1.4126324, 0.96926431)\},
\cr
\De_{\ell=2}^{\rm gap} \in \{
& 3, 4, 5
\}.
}}
Here, $\De_{\ell=2}^{\rm gap}$ is a lower bound on the dimension of $\Z_2$-even spin-2 operators, other than the stress tensor $T_{\mu\nu}$. When $\De_{\ell=2}^{\rm gap}>3$, we additionally impose the Ward identity $C_{\s\s T}/C_{\e\e T}=\De_\s/\De_\e$.  As in \refs{\KosBKA,\SimmonsDuffinQMA,\ONFuture}, we also impose that $\s,\e$ are the only relevant scalars. The number of derivatives of the crossing equation used is $\Lambda=35$, in the notation of \SimmonsDuffinQMA. The choices in \spectrumchoices\ are somewhat arbitrary. It would be interesting to do a more systematic scan over the full allowed region in \ONFuture.

The first few ``stable" operators in the $\e\times\e$, $\s\times\s$, and $\e\times\s$ OPEs are summarized in Tables 1 and 2. 

\bigskip

\vbox{
\offinterlineskip
\lineskip=3pt
\noindent\centerline{\vbox{\tabskip=0pt \offinterlineskip
\hrule
\halign{&\vrule# &\strut \ \hfil#\  \cr
& $\cO$ \hfil &&  $\Z_2$  \hfil   &&  $\ell$ \hfil && $\Delta$  \hfil && $C_{\s\s\cO}$\hfil  && $C_{\e\e\cO}$ \hfil &\cr
\noalign{\hrule}
& $\e$ \hfil && $+$  \hfil &&  0  \hfil   &&   $1.412625(10)$ \hfil &&  $1.0518537(41)$ \hfil && $1.532435(19)$\hfil   &\cr
& $\e'$ \hfil && $+$  \hfil &&  0  \hfil   &&   $3.82966(9^*)$ \hfil && $0.053029(3^*)$\hfil &&  $1.53643(3^*)$ \hfil  &\cr
& $T_{\mu\nu}$ \hfil && $+$ \hfil && 2 \hfil && 3 \hfil && $\Delta_\sigma/\sqrt{c}$\ $(c=0.946539(1))$ \hfil && $\Delta_\epsilon/\sqrt{c}$ \hfil &\cr
& $T'_{\mu\nu}$ \hfil && $+$ \hfil && 2 \hfil && $5.509(1^*)$ \hfil && $0.0172(2^*)$ \hfil && $1.12(1^*)$ \hfil &\cr
& $C_{\mu\nu\rho\s}$ \hfil && $+$ \hfil && 4 \hfil && $5.02274(4^*)$ \hfil && $0.1319(1^*)$ \hfil && $0.4731(4^*)$ \hfil &\cr
}
\hrule}
}
\vskip0.2cm
\noindent{{\bf Table 1:} {\it $\Z_2$-even operators in the 3d Ising model with dimension less than $6$. Non-rigorous errors are marked with stars. (They are given by the standard deviation of solutions to crossing symmetry described in the text.) Other errors are rigorous.}}
}

\bigskip

\vbox{
\offinterlineskip
\lineskip=3pt
\noindent\centerline{\vbox{\tabskip=0pt \offinterlineskip
\hrule
\halign{&\vrule# &\strut \ \hfil#\  \cr
& $\cO$ \hfil &&  $\Z_2$  \hfil   &&  $\ell$ \hfil && $\Delta$  \hfil && $C_{\s\e\cO}$\hfil &\cr
\noalign{\hrule}
& $\s$ \hfil && $-$  \hfil &&  0  \hfil   &&   $0.518149(1)$ \hfil && $1.0518537(41)$ \hfil  &\cr
& $\s'$ \hfil && $-$  \hfil &&  0  \hfil   &&   $5.292(1^*)$ \hfil && $0.05722(3^*)$ \hfil  &\cr
&  \hfil && $-$  \hfil &&  2  \hfil   &&   $4.18031(1^*)$ \hfil && $0.635494(1^*)$ \hfil  &\cr
&  \hfil && $-$  \hfil &&  3  \hfil   &&   $4.6378(8^*)$ \hfil && $0.249(3^*)$ \hfil  &\cr
}
\hrule}
}
\vskip0.2cm
\noindent{{\bf Table 2:} {\it $\Z_2$-odd operators in the 3d Ising model with dimension less than 6. Non-rigorous errors are marked with stars, as in Table~1.}}
}

\bigskip

We have the conformal block expansions
\eqn\conformalblockexpansions{\eqalign{
g_{\e\e\e\e}(u,v) &=
\sum_{\cO\in \e\times\e} C_{\e\e\cO}^2 g^{0,0}_{\Delta,\ell}(u,v),\cr
g_{\s\e\e\s}(u,v) &=
\sum_{\cO\in \s\times\e} C_{\s\e\cO}^2 (-1)^{\ell} g^{\Delta_\s-\Delta_\e,\Delta_\e-\Delta_\s}_{\Delta,\ell}(u,v),\cr
g_{\s\e\e\s}(v,u) &= \frac{v^{\frac{\Delta_\e+\Delta_\s}{2}}}{u^{\Delta_\e}} g_{\s\e\e\s}(u,v) = \sum_{\cO\in \s\times \s} C_{\e\e\cO}C_{\s\s\cO} g^{0,0}_{\Delta,\ell}(u,v),
}}
where $\Delta=\Delta_\cO$, $\ell=\ell_\cO$ for brevity.  The last line gives an expansion for $g_{\s\e\e\s}(v,u)$ that converges at small $u$, which is needed for the second term in \gammams.  We compute the conformal blocks $g_{\Delta,\ell}^{\Delta_{12},\Delta_{34}}(u,v)$ in an expansion in $\rho,\bar \rho$ up to order $\rho^{10}$ using the recursion relation of \KosBKA.\foot{Our conventions are such that $g_{\Delta,\ell}^{\Delta_{12},\Delta_{34}}(u,v)\approx \frac{\ell!}{(2\nu)_\ell}(-1)^\ell C_\ell^{\nu}(\cos\theta)(4r)^\Delta + O(r^{\Delta+1})$, where $\rho=re^{i\theta}$, $\nu=\frac{d-2}{2}$ and $C_\ell^\nu(\cos\theta)$ is a Gegenbauer polynomial. This differs from the conventions of \KosBKA\ by $g_{\Delta,\ell}^{\rm here} = 4^\Delta g_{\Delta,\ell}^{\rm there}$.}

The operators listed in table 2 are enough to compute the two-loop coefficient $\gamma_2$ to one part in $10^{-3}$. Plugging them into the expansions \conformalblockexpansions, we can compute the integral \gammams\ as described in appendix C to obtain
\eqn\twoloopspin{\eqalign{
\gamma_{\cS,2} &= 4.913 -\frac{120.5}{n}.
}}
From \fixeddis, the fixed-point value of the coupling is
\eqn\fixedpointcoupling{
g_* = \frac{\sqrt{2n(n-1)}}{2S_{d-1}}\frac{\de}{2-\frac 1 2 C_{\e\e\e}^2},
}
so that
\eqn\deformedspin{\eqalign{
\De_{\cS}^{\rm IR} &= \De_\s + \gamma_{\cS,2}g_*^2
\approx 0.535-0.0178 n+0.000697 n^2.
}}
Setting $n=0$, we obtain a two-loop estimate for the dimension of the spin operator in the disordered theory
\eqn\deformedspindim{\eqalign{
\Delta_{\cS}^{\rm IR}(n=0) \approx 0.535.
}}
The result \deformedspindim\ is in agreement with experimental results based on the critical exponent $\gamma$, $0.550(25)$ \Belanger, but slightly in tension with results for the ratio $\beta/\nu$ in experiment, $0.507(15)$ \Belanger, and Monte Carlo, $0.505(23)$ \MC.

Although the results are scheme dependent, it is nonetheless interesting to apply formulae from minimal subtraction to compute $\b_2,\gamma_{\cE,2}$ in the 3d Ising CFT.  Plugging the expressions for $\<\cO\cO\cO\cO\>$ and $\<\cE\cO\cO\cE\>$ in appendix~B into \betatwoms, \gammams, and using the conformal block expansion, we obtain
\eqn\twoloopcoefficients{\eqalign{
\beta_2 &=\frac{14.09 n^2-1687 n+3233}{n(n-1)},\cr
\gamma_{\cE,2} &= 22.54 -\frac{555.3}{n},\cr
}}
In minimal subtraction, the two-loop IR dimensions of operators are given by
\eqn\deformeddimensions{
\Delta_{\Phi}^{\rm IR} = \frac{\gamma_{\Phi,1}}{\beta_1}\delta + \frac{\beta_1 \gamma_{\Phi,2}-\beta_2 \gamma_{\Phi,1}}{\beta_1^3}\delta^2 + \dots.
}
Plugging in the above gives
\eqn\deformeddimensionsising{\eqalign{
\Delta_{\cE}^{\rm IR}(n=0) &= \Delta_\e + 1.21092\,\delta - 15.060\,\delta^2 + \dots.\cr
}
}
The two-loop correction is numerically large --- perhaps unsurprising since it is contaminated with lower-loop quantities due to our inability to give a scheme-independent definition.  However, it is interesting to note that the two loop correction has the right sign to push the one-loop result \oneloopalphaestimate\ closer to experimental determinations.

\newsec{Discussion}

We have given two novel ways to understand the random-bond Ising model.  The first is a controlled expansion in $d=2+\varepsilon$ spacetime dimensions.  This has the advantage of being systematically extendable up to arbitrary loop level. At each loop level, we need more information about the continuation of Wilson-Fisher critical exponents in $\varepsilon$. It would be extremely interesting to develop the $(2+\varepsilon)$-expansion directly around the 2d Ising model. (It is understood how to do this for the $O(N)$ models with $N>2$~\BrezinQA, but, as far as we know, for the 2d Ising model it is still not known whether this can be done.)

The second method is an uncontrolled expansion in $3-2\De_\e\approx 0.18$ in the 3d Ising CFT. With this approach, we can leverage precise conformal bootstrap calculations of 3d Ising CFT data. However, we cannot perform a controlled expansion to higher orders within perturbation theory. Nevertheless, the results are encouraging, giving good agreement with experimental determinations of both $\De_\e^{\rm dis}$ and $\De_\s^{\rm dis}$.

Both of these methods have significant advantages over the $\sqrt{\widetilde \epsilon}$-expansion around 4 dimensions. For example, both give the correct sign for the disordered head-capacity at leading order, whereas the $\sqrt{\widetilde \epsilon}$ expansion gives the incorrect sign until 5 loops.
 
 There is a third, controlled, method that we have not yet explored in detail. It is interesting to observe that in the pure $d=3$ $O(2)$ model the dimension of the energy operator is given by $\Delta_\epsilon^{O(2)}\sim 1.51$ and therefore the random bond disorder is only very slightly marginally irrelevant. Logarithmic corrections to the heat capacity would therefore dominate for small but not too small disorder. Let us quote the beta function for disorder in this model. Using~\betagain\ we find
 \eqn\Otwo{{dc^2\over d\log(\mu)} = 0.02 c^2+S_2\left(2-{1\over 2}(C^{O(2)}_{\e\e\e})^2\right)c^4+O(c^6).}
The value of the OPE coefficient $C_{\e\e\e}^{O(2)}$ can be determined with the methods of~\ONFuture. (Alternatively, it can be estimated using the epsilon expansion or the large $N$ expansion.) One finds 
$C_{\e\e\e}^{O(2)}\sim 0.83$. The one-loop correction therefore has a positive sign and it leads to a logarithmic decrease in the effective disorder as long as $0.01 \ll S_{2}c^2\ll1$. It would be nice to observe this effect, which is in essence similar to the logarithmic behavior in the 2d Ising model with random bonds. 

If we now imagine that we can study the $O(N)$ models with continuous $N$, we see that for some $N_c\in (1,2)$ (with $N_c$ very close to $2$ from below) the random bond disorder becomes genuinely marginal. One can have a controlled expansion  in $N-N_c$ and eventually try to study the $N=1$ model in this way.

It is interesting to consider whether the disordered Ising model can be bootstrapped directly. Firstly, one can ask whether the disordered theory has conformal symmetry. This would follow if the deformed $n$-copy Ising model were conformally-symmetric for each $n$ (as long as the analytic continuation to $n=0$ does not destroy the conformal Ward identities) and if the limit $n\rightarrow 0$ commutes with the infrared limit~\AharonyAEA.

The $2+\varepsilon$ expansion shows that conformal symmetry holds perturbatively in $\varepsilon$, and we suspect it is likely to hold nonperturbatively. Another question is whether the theory is reflection positive. Because it is obtained by analytic continuation in $n$, there is no particular reason to expect reflection positivity.\foot{For example, reflection positivity fails in the Wilson-Fisher theory for non-integer $\e$ \HogervorstAKT\ and in $O(N)$ models for non-positive integer $N$ \MaldacenaJN.}  On the other hand, the exclusion plot in \KosBKA\ does not immediately apply to the dimensions $(\De^{\rm IR}_\cS, \De^{\rm IR}_\cE)$, since the disordered theory has the additional relevant operator $\sum_{A\neq B}\s_A \s_B$ which can appear in the $\cS\times \cS$ OPE.\foot{This operator is not turned on in the RG flow by $\sum_{A\neq B}\e_A\e_B$ because it is protected by the $\Z_2^n$ global symmetry of the replicated theory. It would be interesting to estimate its dimension and compare with bootstrap bounds.}  Perhaps the severe truncation method \GliozziYSA, which does not rely on reflection positivity, could shed light on the disordered theory.
 
 Finally, the methods developed here could be used in a variety of other situations. For example, it is curious that a random boundary magnetic field in the $d=2$ Ising model is marginal. It turns out that it is marginally irrelevant~\refs{\BoundaryRGC,\BoundaryRG}. This can be used as a starting point for various interesting calculations 
in conformal perturbation theory that we plan to consider in the future. Another example is the random symmetric rank-two tensor anisotropy in the XY model that one can attempt to study with our methods. In $d=3$ this kind of disorder is relevant and it presumably drives the system to a new fixed point. This is in contrast with the isotropic random-bond disorder in the $d=3$ XY model,~\Otwo, which is slightly irrelevant.

An outstanding open problem is that of disorder in quantum systems. In that case one is instructed to consider a time-independent random source. The replica trick allows to convert this into a non-local-in-time interaction which may be formally relevant, marginal, or irrelevant. The renormalisation of such non-local Hamiltonians should be studied with care. One could hope that a systematic conformal perturbation theory can be developed.

\bigskip
\noindent{\bf Acknowledgments}

We thank Amnon Aharony, Ofer Aharony, Leon Balents, Sean Hartnoll, Luca Iliesiu, Mehran Kardar, Igor Klebanov, Elias Kiritsis, Leonid Levitov, Andreas Ludwig, David Poland, Silviu Pufu, Leonardo Rastelli, Slava Rychkov,  Tadashi Takayanagi, and Shimon Yankielowicz for useful discussions. We are especially grateful to John Cardy for essential comments and collaboration in the early stages of this work. DSD is supported by DOE grant number DE-SC0009988 and a William D. Loughlin Membership at the Institute for Advanced Study.
ZK is supported in part by an Israel Science Foundation center for excellence grant
and by the I-CORE program of the Planning and Budgeting Committee and the Israel Science Foundation (grant number 1937/12). ZK is also supported by the ERC STG grant 335182 and by the United States-Israel Bi-national Science Foundation (BSF) under grant 2010/629.

\appendix{A}{Double Trace Deformations of Generalized Free Fields}

Consider the following toy example. We assume there exists a generalized free field, $\Psi$,  of dimension $d/2$ and we deform the CFT by $f\int {1\over \sqrt 2}\Psi^2$. We normalize $\Psi$ such that $\Psi(x)\Psi(0)\sim {1\over x^d}+\dots$. Therefore, the operator $\cO\equiv {1\over \sqrt 2} \Psi^2$ has a unit-normalized two-point function. The OPE coefficient $C_{\cO\cO\cO}$ is given by \eqn\mftopecoeff{\cO(x) \cO(0) \sim {2\sqrt 2\over x^d} \cO(0)\quad \Longrightarrow \quad C_{\cO\cO\cO}=2\sqrt 2.}
The connected  four-point function is 
\eqn\connectedouble{F(z)=\left\langle \cO(0)\cO(x)\cO(e_1)\cO(\infty) \right\rangle_c = 4  \left( {1\over x^d (x-e_1)^d} +{1\over x^d}+{1\over (x-e_1)^d} \right).  }
In this theory, the beta function is one-loop exact. It is given by $\beta_f=-\sqrt 2f^2 $.

The integrated connected four-point function is thus as in~\formi,
\eqn\dtwol{4\int {d\rho\over \rho} \int d^d x \left({1\over x^d (x-e_1)^d} +{1\over x^d}+{1\over (x-e_1)^d}\right).} 
The main point is that we know that in any physical scheme~\dtwol\ does not produce a single log because the beta function is one-loop exact. Therefore we can subtract this term to cancel all the double logs in~\formi. We therefore find~\betatwo.

It is also interesting to consider the anomalous dimension of $\Psi$. In this computation one encounters the four-point function as in~\lambdaexp\
\eqn\heatgenfree{\langle\Psi(0)\cO(e_1)\cO(x)\Psi(\infty)\rangle_c=2\left({1\over x^d(x-e_1)^d}+{1\over (x-e_1)^d}\right).} 
We know that the dimension of $\Psi$ is one-loop exact~\foot{We easily see that
$C_{\Psi\Psi \cO}=\sqrt 2$ and hence $\Delta_\Psi(f)=\Delta_\Psi-\sqrt 2 f$. } and therefore $A'=0$ in~\Udim.
This implies that there is no single-log in the integral 
\eqn\combtwo{\int {d\rho\over \rho} \int d^dz \left( {1\over z^d(z-1)^d}+{1\over (z-1)^d}\right).}
Therefore we can subtract an appropriate 
combination of~\dtwol\ and \combtwo\ 
in order to remove the  $\log^2$ divergences from~\integratedim\ as in~\finalA.

\appendix{B}{Correlators in the Replicated Ising Model}

Here, we collect results for correlators in the $n$-fold tensor product of the Ising Model.  We are interested in the operators
\eqn\replicaoperators{\eqalign{
\cO(x) &\equiv {1 \over \sqrt{2n(n-1)}} \sum_{A\neq B} \epsilon_A(x)\epsilon_B(x),\cr
{\cal S}(x) &\equiv {1 \over \sqrt n} \sum_A \sigma_A(x),\cr
{\cal E}(x) &\equiv {1 \over \sqrt n} \sum_A \epsilon_A(x),
}}
which are defined so that they are canonically normalized when $\sigma$ and $\epsilon$ are.

Correlators in the replicated theory can be reduced to correlators in a single copy of the Ising model via simple combinatorics.
We have the following three-point coefficients (computed in section~5):
\eqn\replicathreept{\eqalign{
C_{\cO\cO\cO} &= {4(n-2)+2C_{\e\e\e}^2 \over \sqrt{2n(n-1)}},\cr
C_{\cE\cE\cO} &= \sqrt{2(n-1) \over n},\cr
C_{\cS\cS\cO} &= 0.
}}

Let us now compute the four-point functions needed in this work.  Let
\eqn\oprime{
\cO'(x) =\sum_{A\neq B} \e_A(x)\e_B(x)
}
be our deformation before normalization.  After some combinatorics, we find
\eqn\oprimefourpt{\eqalign{
&\<\cO'(x_1)\cO'(x_2)\cO'(x_3)\cO'(x_4)\> \cr
&= 8 n(n-1) \< 1 2 3 4\>^2
\cr
&+16 n(n-1)(n-2)\biggr[\< 1 2 3 4\>\bigr(\< 1 2\>  \< 3 4\>  
 +\< 1 3\>  \< 2 4\>   + \< 2 3\> \< 1 4\> \bigr)
 \cr
 &
\qquad+\< 1 3\>  \< 1 2 4\>  \< 2 3 4\> +\< 1 2\>  \< 1 3 4\>  \< 2 3 4\> + \< 2 3\> \< 1 2 4\>  \< 1 3 4\>
\cr
&
\qquad+   \< 3 4\>\< 1 2 3\>\< 1 2 4\>  +\< 2 4\>  \< 1 2 3\>\< 1 3 4\> +\< 1 4\>  \< 1 2 3\>\< 2 3 4\>\biggr] \cr
&
+4 n(n-1)(n-2)(n-3) \biggr[\< 2 3\>^2 \< 1 4\>^2 +\< 1 3\>^2 \< 2 4\>^2+\< 1 2\>^2 \< 3 4\>^2
\cr
&
\qquad+4 \< 1 3\>  \< 2 3\> \< 2 4\> \< 1 4\> +4 \< 1 2\>  \< 2 3\> \< 3 4\> \< 1 4\> 
+4 \< 1 2\>  \< 1 3\>  \< 2 4\>  \< 3 4\> \biggr],
}}
where on the right-hand side, we use shorthand notation for correlators in the pure (non-replicated) theory:
\eqn\shorthand{
\<\e(x_i)\cdots \e(x_j)\> \to \<i\cdots j\>.
}

Replacing the two- and three-point functions by their actual values, we get
\eqn\oprimefourptactual{\eqalign{
&\<\cO'(0)\cO'(x)\cO'(e_1)\cO'(\oo)\> \cr
&= 8n(n-1) \<\e(0)\e(x)\e(e_1)\e(\oo)\>^2 \cr
& + 16n(n-1)(n-2) \<\e(0)\e(x)\e(e_1)\e(\oo)\>
\left(1 + {1 \over x^{2\De_\e}} + {1 \over (x-e_1)^{2\De_\e}}\right)
\cr
& + 32n(n-1)(n-2) {C_{\e\e\e}^2 \over x^{\De_\e}|x-\e_1|^{\De_\e}}
\left(1 + {1 \over x^{\De_\e}} + {1 \over (x-e_1)^{\De_\e}}\right)
\cr
&+4n(n-1)(n-2)(n-3)
\left(
1+{1 \over x^{4\De_\e}} + {1 \over (x-e_1)^{4\De_\e}} 
 \right.
\cr
& 
\left.
\qquad\qquad\qquad\qquad\qquad\qquad+4 \left(
{1 \over x^{2\De_\e}}
+ {1 \over (x-e_1)^{2\De_\e}}
+ {1 \over x^{2\De_\e}(x-e_1)^{2\De_\e}}
\right)
\right).
}}
From here, we obtain the connected four-point function of $\cO'$ in terms of connected four-point functions in the pure theory,
\eqn\oprimeconnected{\eqalign{
&\<\cO'(0)\cO'(x)\cO'(e_1)\cO'(\oo)\>_c \cr
&= 8n(n-1) \<\e(0)\e(x)\e(e_1)\e(\oo)\>_c^2
\cr
&+ 16n(n-1)^2 \<\e(0)\e(x)\e(e_1)\e(\oo)\>_c
\left(1 + {1 \over x^{2\De_\e}} + {1 \over (x-e_1)^{2\De_\e}}\right)
\cr
& + 32n(n-1)(n-2) {C_{\e\e\e}^2 \over x^{\De_\e}|x-\e_1|^{\De_\e}}
\left(1 + {1 \over x^{\De_\e}} + {1 \over (x-e_1)^{\De_\e}}\right)
\cr
&+16n(n-1)(n^2-3n+3)
\left(
{1 \over x^{2\De_\e}} + {1 \over (x-e_1)^{2\De_\e}} + {1 \over x^{2\De_\e}(x-e_1)^{2\De_\e}}
\right).
}
}

When disorder is marginal, $\De_\e=d/2$, the above four-point function has singularities of the form $x^{-d}$ and $x^{-3d/2}$.  To compute the beta function coefficient $\b_2$, we can subtract them off using the prescriptions~\betatwo\ and~\betatwoimp, and then integrate, leading to
\eqn\regulatedoprimeint{\eqalign{
{\cal I}(n)
&\equiv{1 \over 2n(n-1)}\int d^d x \biggr[\<\cO'(0)\cO'(x)\cO'(e_1)\cO'(\oo)\>_c 
\cr
& \qquad\qquad\qquad- n(n-1)(4(n-2)+2C_{\e\e\e}^2)^2 \left({1\over x^d (x-e_1)^d} +{1\over x^d}+{1\over (x-e_1)^d}\right)
\cr
&
\qquad\qquad\qquad-16n(n-1)^2C_{\e\e\e}^2\left({1 \over x^{3d/2}}+{1 \over (x-e_1)^{3d/2}}+x^{d/2}\right)\biggr],
}}
where we have included the additional factor $(2n(n-1))^{-1}$ for later convenience.
Plugging in \oprimeconnected\ with $\De_\e=d/2$ and $F(x)=\<\e(0)\e(x)\e(e_1)\e(\oo)\>_c$ gives equation \replicaeqii\ in the main text.

We will also use the following two formulae 
\eqn\epscombinatorics{\eqalign{
&\left\langle\sum_{A} \epsilon_A(0)\cO'(x)\cO'(e_1)\sum_{B}\epsilon_B(\infty)\right\rangle_c
\cr
&=4n(n-1)\left(\langle \epsilon(0)\epsilon(e_1)\epsilon(x)\epsilon(\infty)\rangle_c\langle\epsilon(e_1)\epsilon(x)\rangle
+\langle \epsilon(0)\epsilon(e_1)\epsilon(x)\rangle \langle \epsilon(e_1)\epsilon(x)\epsilon(\infty)\rangle
\right)
\cr
&
+4n(n-1)^2\left(
\langle \epsilon(0)\epsilon(e_1)\rangle\langle\epsilon(e_1)\epsilon(x)\rangle \langle \epsilon(x)\epsilon(\infty)\rangle
+\langle \epsilon(0)\epsilon(x)\rangle\langle\epsilon(e_1)\epsilon(x)\rangle \langle \epsilon(1)\epsilon(\infty)\rangle
\right)
}}
and
\eqn\sigcombinatorics{
\left\langle\sum_{A} \sigma_A(0)\cO'(x)\cO'(e_1)\sum_{B}\sigma_B(\infty)\right\rangle_c
=4n(n-1)\langle \sigma(0)\epsilon(x)\epsilon(e_1)\sigma(\infty)\rangle_c\langle\epsilon(x)\epsilon(e_1)\rangle.
}

Putting everything together, the relevant four-point functions of canonically normalized operators \replicaoperators\ are given by
\eqn\replicafourpt{\eqalign{
&\<\cO(x_1)\cO(x_2)\cO(x_3)\cO(x_4)\> = {1 \over n(n-1)}{1 \over x_{12}^{4\De_\e} x_{34}^{4\De_\e}} \left[\phantom{{u^1_1 \over v^1_1}}\!\!\!\!\!\!
2 g_{\e\e\e\e}(u,v)^2\right.\cr
&
\left.\qquad
+ 4 (n-2) \left (\left(1+u^{\De_\e }+{u^{\De_\e }\over v^{\De_\e }}\right) g_{\e\e\e\e}(u,v)
+
2C_{\e\e\e}^2 \left({u^{3 \De_\e /2} \over v^{\De_\e /2}}+{u^{3 \De_\e /2} \over v^{\De_\e }}+{u^{\De_\e } \over v^{\De_\e/2}}\right)\right)\right.\cr
&
\left.\qquad
+ (n-2)(n-3) \left(1+u^{2 \De_\e }+{u^{2 \De_\e } \over v^{2 \De_\e }}+4 \left({u^{2 \De_\e } \over v^{\De_\e }}+u^{\De_\e }+ {u^{\De_\e } \over v^{\De_\e}}\right)\right)
\right],\cr
&\<\cE(x_1)\cO(x_2)\cO(x_3)\cE(x_4)\> = \frac{1}{n}\frac{1}{x_{12}^{2 \Delta_\epsilon } x_{23}^{2 \Delta_\epsilon } x_{34}^{2 \Delta_\epsilon }} \cr
&
\qquad
\times \left[
2 g_{\epsilon \epsilon \epsilon \epsilon}(u,v)
+
2 C_{\epsilon \epsilon \epsilon}^2 u^{\Delta_\epsilon /2}
+
(n-2) \left(2+2 u^{\Delta_\epsilon }+\frac{u^{\Delta_\epsilon }}{v^{\Delta_\epsilon }}\right)
\right],\cr
&\<\cS(x_1)\cO(x_2)\cO(x_3)\cS(x_4)\> =
\frac{1}{n}\frac{1}{x_{12}^{2 \Delta_\sigma } x_{34}^{2 \Delta_\sigma }x_{23}^{4 \Delta_\epsilon-2 \Delta_\sigma}}\cr
&
\qquad
 \times \left[2  u^{\frac{\Delta_\sigma -\Delta_\epsilon }{2}}v^{\Delta_\epsilon-\Delta_\sigma  } g_{\sigma \epsilon \epsilon \sigma }(u,v)
  +
  (n-2) \frac{u^{\Delta_\sigma}}{v^{\Delta_\sigma}} \right],
}}
where the Ising model four-point functions are defined by
\eqn\isingfourpt{\eqalign{
\<\e(x_1)\e(x_2)\e(x_3)\e(x_4)\> &= \frac{g_{\e\e\e\e}(u,v)}{x_{12}^{2\De_\e}x_{34}^{2\De_\e}},\cr
\<\s(x_1)\e(x_2)\e(x_3)\s(x_4)\> &= \left(\frac{x_{24} x_{13}}{x_{14}^2}\right)^{\De_\s-\De_\e}\frac{g_{\s\e\e\s}(u,v)}{x_{12}^{\De_\e+\De_\s} x_{34}^{\De_\e+\De_\s}}.
}}

\appendix{C}{Regulated Integration Over ${\cal R}$}

In our two-loop calculations in the minimal subtraction scheme, we must integrate over the region
\eqn\regionR{
{\cal R} = \{x\in \R^3 : |x|\leq 1,|x|\leq |e_1-x|\}.
}
The integral should be regulated in such a way that we throw away power law divergences coming from integrating $|x|^{-a}$ with $a>d=3$ at the origin.  Furthermore, when terms $|x|^{-d+\delta}$ lead to large contributions $\sim\frac 1 \delta$, we must compute the integral with high-enough precision to subtract off the poles.  Our strategy will be to expand the integrand as a sum of terms so that the integral can be performed exactly for each term.  The accuracy of the result will depend on the number of terms we keep.

Our integrands are most naturally expressed as a series in $r$, where each term is a polynomial in $\eta=\cos\theta$, with
\eqn\reta{
r e^{i\theta} = \frac{z}{(1+\sqrt{1-z})^2}.
}
If we truncate the series at order $O(r^k)$, the resulting error will go like $(2-\sqrt 3)^k$, where $2-\sqrt 3$ is the maximum value of $r$ over ${\cal R}$.
Thus, it suffices to integrate individual terms in the expansion,
\eqn\sufficestointegrate{
\int_{\cal R} d^d x\,r^a P(\eta) = 2\pi\int_{-1}^1 d\eta \int_{r<r_*(|\eta|)} dr \frac{64 r^2(1-r^2)(1+r^2-2r\eta)}{(1+r^2+2r\eta)^5}r^a P(\eta),
}
where $r_*(|\eta|)$ is the smaller solution of
\eqn\rstareta{
\frac{1-4r+r^2}{2r} = |\eta|.
}
Expanding the integrand in a power series in $r$, we obtain terms of the form
\eqn\sumofterms{\eqalign{
&\int_{-1}^1 d\eta \int_{r<r_*(|\eta|)} dr\, r^b Q(\eta) = \int_{-1}^1 d\eta \frac{r_*(|\eta|)^{b+1}}{b+1} Q(\eta)\cr
&= -\int^{2-\sqrt 3}_{3-2\sqrt 2} dr_* \frac 1 2 \left(1-\frac 1 {r_*^2}\right)\frac{r_*^{b+1}}{b+1}\left(Q\left(\frac{1-4r_*+r_*^2}{2r_*}\right)+Q\left(-\frac{1-4r_*+r_*^2}{2r_*}\right)\right).
}}
where $Q(\eta)$ is a polynomial.
We finally expand the integrand as a series in $r_*$ and integrate exactly term by term.

The error coming from each expansion goes like $(2-\sqrt 3)^k$.  In practice, we take $k=12$.

\listrefs

\end